\documentclass[11pt]{article}

\usepackage[latin1]{inputenc}
\usepackage{amsmath}
\usepackage{amsfonts}
\usepackage{amssymb}
\usepackage{amsxtra}
\usepackage[margin=3cm]{geometry}
\usepackage{color}
\usepackage{hyperref}
\hypersetup{linktocpage=true}
\usepackage{graphicx}
\usepackage{placeins}
\usepackage{upgreek} 
\usepackage{mathrsfs}
\usepackage{accents}
\usepackage{multirow} 

\numberwithin{equation}{section}
\newcommand{\beq}{\begin{equation}}
\newcommand{\eeq}{\end{equation}}
\def\be {\begin{equation}}
\def\ee {\end{equation}}
\def\bs#1\es{\begin{split}#1\end{split}}
\def\ba#1\ea{\begin{align}#1\end{align}}
\def\bg#1\eg{\begin{gathered}#1\end{gathered}}
\def\bea{\begin{eqnarray}}
\def\eea{\end{eqnarray}}

\def\nn{\nonumber}

\def\a{\alpha}
\def\b{\beta}
\def\c{\chi}
\def\d{\delta}
\def\D{\Delta}
\def\e{\epsilon}

\def\f{\phi}

\def\F{\Phi}
\def\g{\gamma}
\def\G{\Gamma}
\def\h{\eta}
\def\k{\kappa}
\def\l{\lambda}
\def\L{\Lambda}
\def\m{\mu}
\def\n{\nu}
\def\o{\omega}
\def\O{\Omega}
\def\p{\psi}
\def\P{\Psi}
\def\q{\theta}
\def\Q{\Theta}
\def\r{\rho}
\def\s{\sigma}
\def\S{\Sigma}
\def\t{\tau}
\def\x{\xi}
\def\z{\zeta}

\def\bls{\bigg [}
\def\brs{\bigg ]}

\newcommand{\cC}{\mathcal{C}}
\newcommand{\cS}{\mathcal{S}}

\newcommand{\cG}{\mathcal{G}}

\newcommand{\cB}{\mathcal{B}}

\def\cL{\mathcal{L}}
\def\cP{\mathcal{P}}

\def\cM{\mathcal{M}} 
\def\cN{\mathcal{N}}
\def\cV{\mathcal{V}}

\def\bP{\mathbb{P}}
\def\bZ{\mathbb{Z}}

\def\Re{\text{Re}}
\def\Im{\text{Im}}
\def\Tr{\text{Tr}}
\def\ln{\text{ln}}

\def\dim{\text{dim}}
\def\rank{\text{rank}}
\def\const{\text{const}}
\def\sign{ \text{sign} }
\def\flux{ \text{flux} }
\def\Coulomb{ \text{Coulomb} }

\def\pr{\prime} 
 
\def\sq{\sqrt}
\def\pa{\partial}
\def\na{\nabla}
\def\fr{\frac}
\def\id{\rlap 1\mkern4mu{\rm l}}
\def\we{\wedge}
\def\ra{\rightarrow}

\newcommand{\wh}[1]{ {\hat{#1}}{} }
\newcommand{\til}[1]{ {\tilde{#1}} }

\let\foo\bar 
\renewcommand{\bar}[1]{ {\foo{  #1} }{} }

\newlength{\dhatheight}
\newcommand{\whh}[1]{%
    \settoheight{\dhatheight}{\ensuremath{\hat{#1}}}%
    \addtolength{\dhatheight}{-0.35ex}%
    \hat{\vphantom{\rule{1pt}{\dhatheight}}%
    \smash{\hat{#1}}}{}}

\def\evNM{|_{\cN_{(M)}=1}}

\def\fF{\check F}
\def\ff{\check \F}
\def\fV{\check \cV}
\def\ft{\check \t}
\def\bft{\check {\bar{\t}}{}}

\def\fk{{\k^\pr}}
\def\fl{{\l^\pr}}
\def\fm{{\d^\pr}}
\def\fn{{\g^\pr}}

\def\fO{\O}

\def\Tp{ T_{\cB} }

\begin{document}

\baselineskip=16pt
\setlength{\parskip}{6pt}

\begin{titlepage}
\begin{flushright}
\parbox[t]{1.4in}{
\flushright MPP-2013-4}
\end{flushright}

\begin{center}

\vspace*{1.5cm}

{\LARGE \bf Gauged supergravities and their \\[.3cm] symmetry-breaking vacua 
              in F-theory 
}

\vskip 1.5cm

\renewcommand{\thefootnote}{}

\begin{center}
   \bf{Thomas W.~Grimm} and \normalsize \bf{Tom G. Pugh}\footnote{\texttt{grimm, pught @mpp.mpg.de}} 
\end{center}
\vskip 0.5cm

 \emph{ Max Planck Institute for Physics, \\ 
        F\"ohringer Ring 6, 80805 Munich, Germany} 

\end{center}

\vskip 1.5cm
\addtocounter{footnote}{-1}
\renewcommand{\thefootnote}{\arabic{footnote}}

\begin{center} {\bf ABSTRACT } \end{center}
We first derive a class of six-dimensional $(1,0)$ gauged
supergravities arising from threefold compactifications of F-theory
with background fluxes. The derivation proceeds via the M-theory dual
reduction on an $SU(3)$-structure manifold with
four-form $G_4$-flux. We then show that vacuum solutions of these
six-dimensional theories describes four-dimensional flat space times a
compact two-dimensional internal space with additional localized
sources. This induces a spontaneous compactification to four
space-time dimensions and breaks the supersymmetry from $\cN=2$ to $\cN=1$,
which allows the reduced theory to have a
four-dimensional chiral spectrum. We perform the reduction explicitly
and derive the $\cN=1$ characteristic data of the four-dimensional
effective theory. The match with fourfold reductions of F-theory is
discussed and many of the characteristic features are compared. We
comment, in particular, on warping effects
and one-loop Chern-Simons terms generically present in
four-dimensional F-theory reductions.
\end{titlepage}

\newpage
\noindent\rule{15cm}{.1pt}		
\tableofcontents
\vspace{20pt}
\noindent\rule{15cm}{.1pt}

\setcounter{page}{1}
\setlength{\parskip}{9pt} 

\newpage

\section{Introduction}

Recently much effort has focused on the study of F-theory compactifications. This includes both the study 
of F-theory vacua as well as the determination of the supersymmetric 
effective actions \cite{Vafa:1996xn, Denef:2008wq, Weigand:2010wm, Maharana:2012tu}. F-theory is a non-perturbative formulation 
of Type IIB string theory with space-time filling seven-branes, in which  
 the complexified Type IIB string coupling is geometrized in the complex
structure of an auxiliary two-torus. Crucial parts of the seven-brane 
physics can then be captured geometrically by studying degenerations 
of this torus. More recently, it 
was argued that they also can admit appealing realizations
of Grand Unified Theories \cite{Donagi:2008ca, Beasley:2008dc, Marsano:2009ym, Marsano:2009wr, Blumenhagen:2009yv, Chen:2010ts, Mayrhofer:2012zy, arXiv:1302.1854} and thus provide a very geometrical 
approach to phenomenological questions. The derivation of the 
effective actions are crucial both to infer corrections to the duality
and to study phenomenological setups. 

While there has been significant progress 
in the derivation of the leading classical effective action 
the inclusion of corrections predicted by string theory 
is still challenging. One of the obstacles to deriving these corrections
is the fact that there is no low-energy effective action 
of F-theory and one has to take a detour via M-theory 
to infer properties of F-theory vacua and lower-dimensional 
effective actions \cite{Denef:2008wq, Dasgupta:1999ss, Grimm:2010ks}. The limit from M-theory to F-theory is, however,
very non-trivial and still has not been worked out for general four-dimensional
F-theory reductions.

In order to gain some insights into this we can consider instead a reduction of F-theory to six dimensions \cite{Morrison:1996na, Morrison:1996pp, Ferrara:1996wv, Bonetti:2011mw}. Here the increased amount of supersymmetry and the requirement of anomaly cancellation restricts the discussion and simplifies the analysis. For this reason the effective action of F-theory reduced on a Calabi-Yau threefold was recently described in \cite{Ferrara:1996wv, Bonetti:2011mw}. In what follows we will consider generalisations of this reduction in which the M-theory dual is reduced on an SU(3) structure manifold with four-form flux.  We will then understand the F-theory duals of these effects and show how they may modify the vacua of the 6D effective theories. Finlay we will show how these 6D effective theories may be further reduced to four dimensions and compared with  F-theory reductions on Calabi-Yau fourfolds with similar modifications.

We begin our investigation by deriving the 6D effective gauged
supergravity action obtained by reducing F-theory on an elliptically
fibered Calabi-Yau threefold $Y_3$ with background fluxes. These
fluxes will correspond to worldvolume two-form flux located on the
two-cycle $S$ in the base $B_2$ of $Y_3$ wrapped by a seven-brane.  In practice the derivation proceeds by
uplifting M-theory on a Calabi-Yau threefold with a special class of
four-form fluxes $G_4$ for the M-theory three-form potential. The
uplifted 6D supergravity actions admit a gauged shift symmetry of an
axion in the universal hypermultiplet generally present in a Type IIB
reduction to six dimensions.

We will also consider F-theory reductions in which extra massive U(1) symmetries arise. The M-theory duals of these will result from a reduction on an SU(3) structure manifold for which may be considered to be an appropriately small modification of the original Calabi-Yau threefold. Here the harmonic forms of the threefold are supplemented by additional non-harmonic forms for which the deviation from the harmonic constraint is parametrised by a constant similar to the flux parameters in the case described above. The effective theory again involves additional gauged shift symmetries of the hypermultiplet scalars \cite{Micu:2011aa}. 

In certain cases we will see that the derived gauged theories may no longer admit 6D Minkowski solutions.
However, we find that compactifying solutions exist which contain a 4D
Minkowski factor and a compact two space $\wh \cB$. Generally the scalars in the
6D hypermultiplet moduli space have to admit a non-trivial profile over $\wh \cB$ for these solutions to exist.
Concretely we find solutions preserving four supercharges 
with a profile for the axion $\Phi$ and
the volume $\cV$ of the Calabi-Yau threefold for the F-theory reductions with 7-brane flux. Moreover, these solutions also
describe a flux for the 6D gauge fields on the compact space $\wh \cB$. 
These new fluxes then ensure the self-duality of the
total 7-brane flux on the worldvolume $\cS = \wh \cB \times S$. The solutions are sourced by the presence of additional co-dimension-two
localized sources. These sources fill the 4D Minkowski space and are
points on $\wh \cB$. In what follows we will interpret them as additional seven-branes
wrapping the entire base $B_2$. 

Having found vacuum solutions with compact component $\wh \cB$ 
of the 6D gauged supergravity theory we proceed to derive 
the 4D effective theory encoding the dynamics of the 
fluctuations around the 6D backgrounds. The resulting 
4D theory is shown to be a gauged supergravity theory 
consistent with an F-theory reduction to four dimensions. Remarkably the 6D solutions and the resulting four-dimensional
effective action captures many crucial features of
a general 4D F-theory reduction. We find that the conditions 
on the 6D solutions match with certain 
tadpole cancellation conditions, the 6D flux background 
lifts to a self-dual $G_4$ in a fourfold reduction and a non-trivial warp factor is induced. 
When neglecting warping effects the 4D $\cN=1$ characteristic data
are shown to match the results of \cite{Denef:2008wq, Grimm:2010ks, Grimm:2011fx, Grimm:2011tb, Grimm:2011sk}. Similarly we will show that massive U(1) symmetries in the 6D reduction provide the same effects as massive U(1)s in the equivalent 4D F-theory reduction \cite{Grimm:2011tb, Grimm:2010ez}.

Reductions of 6D gauged supergravities to 4-dimensions, on compact spaces similar to those considered here, have been analysied in the past \cite{Salam:1984cj, Aghababaie:2002be,  Aghababaie:2003wz, Aghababaie:2003ar, Burgess:2004kd, Gibbons:2003gp, Pope:2011yi} and higher dimensional origins for these theories have also been proposed \cite{Cvetic:2003xr, Pugh:2010ii}. The reductions we consider here differ from these as the 6D theories we describe have only gauged hypermultiplet shift symmetries, so the fluxes that can be turned on have a different form. However, many of the qualitative effects are comparable. One reason for the interest in these sorts of solutions was based on the idea that local effects at the locations of certain branes in the 6D solutions may provide a natural solution to the cosmological constant problem \cite{Aghababaie:2003wz, Aghababaie:2003ar, Burgess:2004kd}. We will not comment further on this idea here but will briefly mention that in future work it may be interesting to investigate to what degree these mechanisms may be embedded in F-theory by means of an intermediate reduction such as that shown here. 

This paper is organized as follows. In section  \ref{sec:gauged6Dsugras} we will determine the 6D effective theories arising in modified F-theory compactifications by examining the dual M-theory reductions. In Section \ref{sec:6Dvacua} we will examine the vacua of the 6D effective theories and consider the subsequent reduction to 4D. These will then be matched with the effective theories of more direct 4D F-theory compactifications in Section \ref{sec:4Dsugras}. In appendix \ref{Conventions} we will list our conventions. Finally in appendix \ref{DomainWalls} we will describe how certain 6D solutions we have described may be related to the 5D domain wall solutions of \cite{Lukas:1998tt}.

\section{6D gauged supergravity from F-theory and M-theory} \label{sec:gauged6Dsugras}

In this section we derive certain 6D gauged supergravities 
 which may be obtained by reductions of F-theory. These 6D theories are arrived at by taking the F-theory limit 
of a 5D M-theory reduction on an SU(3) structure manifold with 4-form fluxes. 
In Section \ref{M_on_threefold} we briefly 
recall the 5D gauged supergravity action arising as a reduction of M-theory on a Calabi-Yau threefold with $G_4$ flux. We then evaluate the result for the special case of an elliptically 
fibered Calabi-Yau threefold. In Section \ref{M_on_SU(3)} we derive an alternative gauged 5D supergravity which arises by deforming the Calabi-Yau to a SU(3) structure manifold. A general class of 6D gauged 
supergravities is reduced on a circle in Section \ref{6D_on_circle}. 
The 5D actions of Section \ref{M_on_threefold} and Section \ref{M_on_SU(3)} are then matched to the 6D reduction of Section \ref{6D_on_circle} after taking the F-theory limit in Section \ref{Matching5DTheories}. This allows to infer the 
6D actions arising from a reduction of F-theory with either 7-brane fluxes or massive U(1)s.


\subsection{M-theory on Calabi-Yau threefolds with fluxes} \label{M_on_threefold}

We begin by reviewing the reduction of M-theory on a Calabi-Yau threefold  $\hat Y_3$ with $\whh G_4$-flux \cite{Lukas:1998yy, Behrndt:2000zh}. The at lowest order in derivatives the bosonic part of the 11D supergravity action is given by \cite{Cremmer:1978km}
\beq \label{11Daction}
S^{(11)} = \int_{\cM_{11}} \fr12 \whh R \whh * 1 - \fr14 \whh G_4 \we \whh * \whh G_4 
           - \fr1{12} \whh C_3 \we \whh G_4 \we \whh G_4 \, , 
\eeq
where $\whh G_4 = d \whh C_3$ and $\whh R$ is the 11D Ricci scalar for which we use 
the conventions outlined in Appendix~\ref{Conventions}. Here we will indicate 
11D objects by $\whh\ $.
When reducing this action on a Calabi-Yau threefold $\hat Y_3$ we make the following ansatz for  the 11D metric
\ba
d \whh s^2 = g_{m n} d x^m d x^n + 2 g_{\bar \imath j}^{(6)} d\bar y^{\bar \imath} d y^j \, ,
\ea
where $m = 0, \ldots 4$ is a world index on the 5D external space 
which is raised and lowered with the metric $g_{m n}$ 
and $i = 1, \ldots 3$ is a world index on the 
complex threefold which is raised and lowered with 
the metric $g_{\bar \imath j}^{(6)}$. In addition 
we will allow for a background flux $\langle \whh G_4 \rangle = G_4^{\rm flux}$
with indices along $\hat Y_3$. This flux can be expanded in a 
basis of four-forms $\tilde \omega^\L$ representing elements of $H^{4}(\hat Y_3)$ as
\beq \label{exp-G4flux}
  G_4^{\rm flux} = \theta'_\L \tilde \omega^\L\ .
\eeq
The coefficients $\theta'_\L$ are in fact quantized and integral when expanded in 
an integral basis of $H^{4}(\hat Y_3,\mathbb{Z})$.

The real scalars parameterizing the variations of the K\"ahler structure of $\hat Y_3$
are denoted by $v^\L$, while the complex scalars parameterizing the changes in the complex structure of $\hat Y_3$ are
named $z^\k$. Infinitesimally they modify the Calabi-Yau metric by  
\beq \label{CYdeformations}
\d g_{i \bar j}  = -i (\o_\L)_{i \bar j} \, \d v^\L\, ,  \qquad \quad  \d g_{i j} \propto (\bar \c_{\bar \k})_{i \bar k \bar l} \O^{\bar k \bar l}_{\ \ j}\,   \d \bar z^{\bar \k} \, , 
\eeq
where $\L = 1, \ldots h^{1,1} (\hat Y_3)$ and $\k = 1, \ldots h^{1,2}  (\hat Y_3) $. In these expressions we 
have introduced a basis $\o_\L$ of $(1,1)$-forms representing elements of $H^{1,1}(\hat Y_3)$, 
and a basis $\chi_{\kappa}$ of $(2,1)$-forms representing elements of $H^{2,1}(\hat Y_3)$. 
$\O_3$ is the globally defined nowhere-vanishing holomorphic 3-form. Using the forms $\o_\L$
one can also expand the K\"ahler form $J$ of the Calabi-Yau manifold as $J = v^\L \o_{\L}$ defining the
finite $v^\L$. This means that the volume  $\cV = \int_{\hat Y_3} *_6 1$ of $\hat Y_3$ 
is given in terms of $v^\L$ by 
\beq \label{volume_triple_intersect}
 \cV  = \fr1{3!} \int_{\hat Y_3}J \we J\we J = \fr1{3!} \cV_{\L \S \Q} v^\L v^\S v^\Q \, , \qquad \quad \cV_{\L \S \Q} = \int_{\hat Y_3} \o_\L \we \o_\S \we \o_\Q \, , 
\eeq
where $\cV_{\L \S \Q}$ are the triple intersection numbers.
In what follows it is useful to separate off the volume and define the new scalars $L^\L $ which satisfy 
\ba \label{def-cN}
  L^\L= \frac{v^\L}{\cV^{1/3}} \ , \qquad \quad \cN_{(M)}  \equiv  \fr1{3!} \cV_{\L \S \Q} L^\L L^\S L^\Q  = 1 \, . 
\ea
When distributed into five-dimensional supermultiplets the scalars $L^\L$ are part of vector multiplets 
while the volume $\cV$ is part of the universal hypermultiplet.

To complete the reduction one also has to consider fluctuations of the 
M-theory three-form $\whh C_3$. Therefore, we make the ansatz 
\ba
\whh G_4 = d\x^K \wedge \a_K - d \tilde \x_K \wedge \b^K +  F^{\pr \L} \wedge \o_\L + \cG_4 + G^{\rm flux}_4  \, ,
\ea
where $(\x^K,\tilde \x_K )$ are 5D scalars, $F^{\pr \L}=dA^{\pr \L}$ are the
field-strengths of 5D $U(1)$ vectors $A^{\pr \L}$, and $\cG_4=d\cC_3$ is the field strength of 
a 5D three-form $\cC_3$. Here we have introduced a symplectic basis of three-forms $(\a_K,\b^K)$ 
on $\hat Y_3$ representing elements of $H^3(\hat Y_3)$ such that $K=0,\ldots , h^{1,2} (\hat Y_3)$.
For an appropriately chosen basis the only non-vanishing double intersections of the 
$(\a_K,\b^K)$ and $(\omega_\L,\tilde \omega^\L)$ are
\beq \label{intersections_dual}
  \int_{\hat Y_3} \a_K \we \b^L  = \d_K{}^L \, , \qquad \int_{\hat Y_3} \o_\L \we \til \o^\S = \d_\L{}^\S \, . 
\eeq

Let us next turn to the determination of the 5D effective action by 
inserting the reduction ansatz into \eqref{11Daction}. 
Integrating over the Calabi-Yau space and carrying out a Weyl rescaling $g_{m n} \ra \cV^{-\fr23} g_{m n}$ to bring the effective action into the 5D Einstein frame, we find that
\ba \label{5Dreduced_action}
S^{(5)}_{(M)} &= \int_{\cM_5} \bls \fr12 R * 1 - \fr12 G_{\L \S}\, d L^\L \we * d L^\S- \fr12 G_{\L \S} \, F^{\pr \L} \we * F^{\pr \S} - \fr1{12} \cV_{\L \S \Q}\, A^{ \pr \L} \we F^{\pr \S} \we F^{\pr \Q} \nn \\
& - \fr1{ 4 \cV^2} d \cV \we * d \cV - \fr14 \cV^2 \cG_4 \we * \cG_4 - \fr14 ( \x^K d \til \x_K - \til \x_K d \x^K + 2 A^{\pr \L}  \q_{  \L}^\pr) \we \cG_4 -  g_{\k \bar \k} d z^\k \we * d \bar z^{\bar \k} \nn \\
&+ \fr1{4\cV} ( \text{Im} M )^{K L} ( d \til \x_K -  M_{K M} d \x^M )  \we *  ( d \til \x_L -  \bar M_{L N} d \x^N )  - \fr1{8 \cV^2} G^{\L \S} \q^\pr_\L  \q^\pr_\S * 1 \brs \,  , 
\ea
where $M_{K M}(z,\bar z)$ is a complex matrix depending on the scalars $z^\kappa$, and $G_{\L \S}(L)$ is a real matrix depending on the scalars $L^\L$. The inverse of $G_{\L \S}$ is 
denoted by $G^{\L \S}$, while the inverse of $\text{Im} M_{K M}$ is denoted by $(\text{Im} M)^{K M}$.
Explicitly $G_{\L \S}$ is derived to be 
\beq
  G_{\L \S} = \fr{1}{2} \frac{1}{\cV^{1/3}} \int_{\hat Y_3} \o_\L \wedge *_6 \, \o_\S  
        = - \fr12 (\pa_{L^\L} \pa_{L^\S} \, \ln \cN_{(M)}) \evNM \, , 
\eeq
with $\cN_{(M)}$ being the cubic polynomial in $L^\L$ defined in \eqref{def-cN} but evaluated at $1$ only after taking the derivative. 
The explicit expressions for the metric $g_{\kappa \bar \kappa}(z,\bar z)$ and the complex matrix $M_{K M}(z,\bar z)$
can be found in equations \eqref{def-gkappabarkappa} and \eqref{def-MKL} of Appendix \ref{Conventions}. We will not need their precise form in the following.

To bring the action \eqref{5Dreduced_action} into a standard supersymmetric form one 
first has to dualize the three-form $\cC_3$ into a 5D scalar $\Phi$. 
We thus introduce a term in the action which imposes the Bianchi identity for $\cG_4$ given by 
\ba
\D S^{(5)}_{(M)} = \int_{\cM_5} - \fr14 d \F \we \cG_4 \, . 
\label{5DFDualAct}
\ea
Upon varying the action with respect to $\cG_4$, now treated as a fundamental field, we find the equation
\ba
2 \cV^2 *_5 \cG_4 + d \F + 2 A^{\pr \L}  \q^\pr_\L +   \x^K d \til \x_K - \til \x_K d \x^K=0 \, .
\ea
Substituting this back into the effective action \eqref{5Dreduced_action} gives the effective action with $\cG_4$ dualized. At this point it is useful to make a redefinition $\F \ra \F + \til \x_K \x^K$ in order to move into a basis where the scalar $\til \x_K$ is purely axionic, which will be important for comparison with what follows. This gives the 5D effective action
\ba \label{MReducedAction2}
S^{(5)}_{(M)} &= \int_{\cM_5} \bls \fr12 R * 1 - \fr12 G_{\L \S} \, d L^\L \we * d L^\S- \fr12 G_{\L \S} \, F^{\pr \L} \we *  F^{\pr \S} - \fr1{12} \cV_{\L \S \Q}\, A^{\pr \L} \we F^{\pr \S} \we F^{\pr \Q} \nn \\
& - \fr1{ 4 \cV^2} d \cV \we * d \cV -\fr1{16 \cV^2} (D\F +  2\x^K d \til \x_K ) \we * (D\Phi +  2 \x^K d \til \x_K  )  \nn \\ 
&- g_{\k \bar \k} d z^\k \we * d \bar z^{\bar \k} + \fr1{4\cV}  ( \text{Im} M )^{K L}  ( d \til \x_K - \bar M_{K M} d \x^M )\we *  ( d \til \x_L - M_{L N} d \x^N ) - V^{(5)}_{\flux} * 1 \brs \, ,
\ea
where we have abbreviated the invariant derivative $D\Phi$ and the scalar potential $V^{(5)}_{\flux}$ as
\ba
     D\Phi& =d \F + 2 A^{\pr \L}  \q^\pr_\L \, , &  V^{(5)}_{\flux}  &= \fr1{8\cV^2} G^{\L \S} \q^\pr_\L \q^\pr_\S \ .
\ea
These gaugings and the potential they induce then describe the deformation away from the ungauged 5D supergravity caused by the background flux $G^{\rm flux}_4$. 

The M-theory/F-theory duality, which we wish to use in order to lift this 5D action in the F-theory limit, only applies 
when the Calabi-Yau manifold $\hat Y_3$ is an elliptic fibration. More precisely, $\hat Y_3$ can 
be the resolution of a singular elliptic fibration over some base twofold $B_2$. 
When such spaces are considered the divisors of $\hat Y_3$ can be split up into three sets with different origins. 
Here we will label $\o_0$ as the duals of the divisor associated to the section of the elliptic fibration, 
$\o_\a$ are the duals of the divisors associated with divisors of the base, 
and $\o_i$ are the duals of the divisors associated with the resolution
of the singularities of the elliptic fibration. 
The 5D vector multiplets are then similarly split so that the vectors are decomposed as $ A^{\pr \L} = ( A^{\pr 0},  A^{\pr \a} ,  A^{\pr i} )$ and the scalars as $L^\L = (R, L^\a, L^i)$. The intersection numbers $\cV_{\L \S \Q}$ also become constrained such that 
\ba
\cV_{000} &= \O_{\a \b} a^\a a^\b \, , &  \cV_{00\a} &= \O_{\a\b} a^\b & V_{0 \a \b} & = \O_{\a\b} \, , \nn \\
\cV_{\a \b \g} & = 0 \, , & \cV_{0 i \L} & = 0 \, , & \cV _{\a \b i} & = 0\, , \nn \\
\cV_{\a ij} & = - C_{ij} \O_{\a \b} b^{\b} \, , & \cV_{ijk} &\neq 0 \, ,  
\label{Vdecomp}
\ea
where $C_{ij}$ is the Cartan matrix of the group associated with the 
singularity resolution of the Calabi-Yau manifold. 

To shift to a basis in which we can lift up to a 6D theory F-theory 
reduction it is helpful to make the following field redefinitions 
\ba  \label{LtoMredef}
M^0 & = 2 R \, ,  & M^\a &= \fr12 L^\a + \fr14 K^\a R  \, , & M^i & = \fr12 L^i \, , \nn \\
A^0 & = 2 A^{\pr 0}  \, , & A^\a &= \fr12  A^{\pr \a} + \fr14 K^\a A^{\pr 0} \, ,  & A^i & = \fr12  A^{\pr i} \, ,\nn \\
\q_0 &= \fr12 \q^\pr _0 - \fr14 K^\a  \q^\pr_\a \, , & \q_\a &= 2  \q^\pr_\a \, , & \q_i &= 2 \q^\pr_i \, . 
\ea
In terms of these redefined fields the scalar $\cN_{(M)}$ then takes the form
\ba \label{DefcNM}
\cN_{(M)} & \equiv  \O_{\a \b} M^0 M^\a M^\b - 4 \O_{\a \b} b^\a C_{ij}  M^{\b} M^i M^j  + \fr1{192} \O_{\a\b} a^\a a^\b M^0 M^0 M^0 \nn \\
&+ \fr12 \O_{\a \b} b^\a C_{ij}  M^0 K^{\b} M^i M^j + \fr43 \cV_{ijk} M^i M^j M^k = 1\, .
\ea
The fields can be arranged into multiplets of the 5D supersymmetry. 
For example the 5D metric $g_{m n}$ together with one of the vectors $A_m^{ 0}$ form 
the bosonic part of the 5D gravity multiplet. The remaining  $h^{1,1}  ( \hat Y_3) - 1$ 
vectors combine with the constrained scalars $L^\L$ to form $n_V^5 = h^{1,1}  ( \hat Y_3) - 1$ 
vector multiplets.  Finally, we note that the $4(h^{1,2}  ( \hat Y_3) + 1)$ scalars given 
by $q^u =  (\cV, \F, z^k, \bar z^{\bar k} , \x^K, \tilde \x_K )$ belong to $n_H^5 = h^{1,2}  (\hat Y_3) +1$ hypermultiplets. 
The resulting 5D action is then given by 
\ba\label{MReducedAction3}
S^{(5)}_{(M)} =\int_{\cM_5} \bls &  \fr12 R * 1 - \fr12 G_{\L \S}  d M^\L \we * d M^\S - \fr12 h_{uv} D q^u \we * D q^v  \\
&- \fr12 G_{\L \S}  F^{\L} \we * F^\S - \fr1{12} \cN_{\L \S \Q} A^\L \we F^\S \we F^\Q - V^{(5)}_{\flux} * 1\brs \, ,\nn
\ea
where $h_{uv}$ is the hypermultiplet target space metric which can be read off by comparison with \eqref{MReducedAction2} and
\beq
G_{\L \S} (M) = - \fr12  (\pa_{M^\L} \pa_{M^\S} \ln \, \cN_{(M)} )\evNM\ ,\qquad    \cN_{\L \S \Q} = (\pa_{M^\L} \pa_{M^\S} \pa_{M^\Q} \cN_{(M)}) \evNM  \, . 
\eeq
In this alternative basis the gauge invariant derivatives and the scalar potential are now given by 
\ba
D q^u &= \left\{ \begin{array}{cc} 
d \F + 2 A^\L \q_\L & \text{ if } q^u = \F \, , \\
d q^u & \text{ if } q^u \neq \F \, , 
\end{array} \right.  & 
 V^{(5)}_{\flux} &= \fr1{8\cV^2} G^{\L \S}\q_\L \q_\S \, . 
 \label{MGaugings1andV_flux}
\ea

In general the potential of a 5D N=1 theory is given by \cite{Andrianopoli:1996vr, Lukas:1998tt} 
\ba
V^{(5)} = - 4 ( G^{\L \S} - 2 M^\L M^\S ) P_\L{}_{A}{}^B P_\S{}_B{}^A + \fr12 h_{uv} k^u_\L k^u_\S M^\L M^\S \,, 
\label{General5DPotential}
\ea
where $k_\L^u$ are the killing vectors which define the gaugings as $D q^u = d q^u + k_\L^u A^\L$ and $P_\L{}_A{}^B$ is a function of the hypermultiplet degrees of freedom, valued in the adjoint of $SU(2)$ and is related to the $SU(2)$ part of the hypermultiplet curvature $K_{u v}$ by
\ba
k^u_\L K_{u v}{}_A{}^B = \na_v P_\L{}_A{}^B \, . 
\ea
The potential found in M-theory reduction we have carried out here results from a special case of this in which \eqref{General5DPotential} becomes simplified as 
\ba
P_{\L}{}_A{}^B P_\S{}_B{}^A & = - \fr1{16} k_\L^u k_\S^v h_{uv} \, . 
\ea
We may then chose a gauge in which this is satisfied as 
\ba
P_\L{}_A{}^B & = \fr{i}{8 \cV} \q_\L \s^3{}_A{}^B \, , 
\label{P3forFlux}
\ea
where $\s^3$ is the Pauli matrix. Substituting this back into \eqref{General5DPotential} then results in the potential  \eqref{MGaugings1andV_flux}.

Let us note here that only shift symmetries are gauged by turning on the flux $G_4^{\rm flux}$. In the M-theory reduction on the resolved $\hat Y_3$ 
there is no charged 5D matter in the effective theory and all gauge fields are $U(1)$ fields. This can be attributed to the 
fact that this 5D theory corresponds to an 6D F-theory compactification on an extra circle when pushing the theory to 
the 5D Coulomb branch.

\subsection{M-theory on $SU(3)$ structure sixfolds} \label{M_on_SU(3)}

In addition to turning on the flux as described above we may also consider 
reductions on a more general class of real six manifolds $\hat Z_6$ that are no
longer Calabi-Yau manifolds \cite{Koerber:2010bx, Samtleben:2008pe, Gurrieri:2002wz, Grana:2006hr, Lopes Cardoso:2002hd}. Concretely we 
will consider in the following six manifolds $\hat Z_6$ that admit $SU(3)$ structure
but which are in general neither K\"{a}hler, nor complex, and do not have vanishing Ricci curvature. However, as a result of the $SU(3)$ structure they do admit a globally defined, no-where vanishing two-form $J$ and three-form $\Omega$.
In contrast to Calabi-Yau spaces with harmonic $J$, $\Omega$ one now has
\ba
  d J &\neq 0 \, , & d\Omega & \neq 0 \, ,
\ea
while we still impose
\beq
  dJ \wedge J = 0  \ . 
\eeq

To perform the reduction we must then expand in a basis that includes both the harmonic forms that we considered before and also a different set of non-closed and exact forms. To avoid extensive notation we will use the 
same indices as in Section \ref{M_on_threefold} and will extend the range of $\L$ and $K$ to include the non-harmonic forms. These then satisfy 
\beq \label{non-closed_forms}
d \a_K  =  e'_{K \L} \til \o^\L\, ,\qquad   d \b^K  = 0\, ,\qquad  d \o_\L  =  e'_{K\L} \b^K\, ,\qquad  d \til \o^\L  = 0 \ .
\eeq
The deviation from Calabi-Yau condition is then described by the constants $e'_{K \L }$. 
These deviations are introduced such that the expanded basis preserves the form of the intersection conditions  \eqref{volume_triple_intersect} and \eqref{intersections_dual} now integrated over $\hat Z_6$. Moreover, we restrict to the case that 
\beq
     \omega_\L \wedge \b^K = 0 \ ,  
\eeq
at least in all integrals. This mimics the conditions valid in Calabi-Yau reductions and accounts 
for the fact that no one-forms are used in the reduction ansatz. 

The dimensional reduction of M-theory on $\hat Z_6$ is performed in analogy with Section \ref{M_on_threefold}
but taking into account the properties \eqref{non-closed_forms} of the forms. For simplicity we will include the flux $G_4^{\rm flux}$
only at the end of the discussion.
The expansion of the M-theory three-form then takes the form 
\beq
\whh G_4  = d \x^K \a_K - D \til \x_K \b^K + F'^\L \o_\L  + \cG_4 + \x^K  e'_{K\L} \til \o^\L \ ,
\eeq
where 
\beq
D \til \x_K =  d \til \x_K +  e'_{K \L}  A'^\L\ . 
\eeq
In order to perform the F-theory lift it will again be necessary to split the index $\L$ into directions associated to the divisors of different origins. In doing this we now extended the range of the index $i$ appearing the the decomposition in order to include the additional non-harmonic 2-forms in \eqref{non-closed_forms}. This means that when making the basis change \eqref{LtoMredef} we may then define
\beq
e_{K 0} = 0 \, , \qquad \quad  e_{K \a} = 0  \, ,\qquad \quad e_{Ki} = 2  e'_{Ki}\ .
\eeq
When carrying out this decomposition we will also extend the definition of $C_{i j}$ appearing in \eqref{Vdecomp} so that now only the part associated with the harmonic 2-forms corresponds to the Cartan matrix of the gauge group, associated with the singularity resolution. Reducing as before, carrying out the rescalings and dualizing the three-from with field strength $\cG_4$ into a scalar $\Phi$ 
we find that
\ba
S^{(5)}_{(M)} = \int_{\cM_5} \bls &\fr12 R * 1 - \fr12 G_{\L \S}\, d M^\L \we * d M^\S - \fr12 h_{uv} D q^u \we * D q^v \nn \\
&- \fr12 G_{\L \S}  \, F^{\L} \we * F^\S - \fr1{12} \cN_{\L \S \Q}\, A^\L \we F^\S \we F^\Q - V^{(5)}_{\rm geom} * 1\brs \ ,
\label{MReducedAction4}
\ea
where $G_{\L \S}(M)$ is formally obtained by the same generating function $\cN_{(M)}$ as in \eqref{DefcNM}. The gaugings that appear here are now given by  
\ba
D q^u = \left\{ \begin{array}{cl} 
d \F +  A^\L e_{K \L} \x^K \, , \quad & \text{ if } q^u = \F \, , \\
d \til \x_K +  A^\L e_{K \L} \, , & \text{ if } q^u = \til \x_K \, , \\
d q^u \, , & \text{ if } q^u \neq \F , \til \x_K \, . 
\end{array} \right.
\label{MGaugings2}
\ea
These can be brought into a simplified form by once again making a field redefinition $\F \ra \F +  \x^K \til \x_K$ which modifies the hypermultiplet metric to match that shown in \eqref{MReducedAction2}. When this is done the scalar $\til \x_K$ has a standard gauged shift symmetry and is the only scalar with a gauge covariantized derivative such that $D \til \x_K  = d \til \x_K +  A^\L e_{K\L}$.

The potential $V^{(5)}_{\rm geom}$ now contains contributions which arise in the M-theory reduction from both the $\whh G_4$ kinetic term and from the internal space Ricci scalar. These combine to give a total potential which agrees with that which is required by supersymmetry \eqref{General5DPotential} for the gaugings we have described. The $SU(2)$ adjoint valued functions $P_\L{}_A{}^B$ can also be derived by reducing the 11D gravitino variation and reading off the relevant term as described in \cite{Gurrieri:2002wz, Grana:2005ny}. For both the fluxes and the geometric deformations we have described here this gives 
\ba
v^\L \cP^3_\L & = \fr{i}{ 8 \cV }  \int_{\hat Z_6} J \we G_4 \, , & v^\L \cP^1_\L + i v^\L \cP^2_\L& =   \fr{i}{8 \sq \cV} e^{\fr12 K_c} \int_{\hat Z_6}  \O \we dJ\, , 
\label{Pcalculation}
\ea
where $K_c$ is the K\"ahler potential for the complex structure deformations $z^\k$ and we have expanded $P_\L{}_A{}^B$ in terms of the Pauli matrices as 
\ba
P_\L{}_A{}^B & = P_\L^x \s^x{}_A{}^B \, , 
\ea
for $x=1,2,3$ . We note that for the SU(3) structure reductions we have considered this gives
\ba
P_\L{}_A{}^B & =  \fr{i e^{\fr12 K_c} }{16 \sq \cV} e_{K \L} ( Z^K + \bar Z^K) \s^1{}_A{}^B +  \fr{e^{\fr12 K_c} }{16 \sq \cV} e_{K \L} ( Z^K -\bar Z^K) \s^2{}_A{}^B + \fr{i}{8 \cV} e_{K \L} \x^K \s^3{}_A{}^B \, ,
\ea
where $Z^K$ are the scalars that appear in the expansion of $\O$ such that we may chose a basis in which $Z^K = \{ 1 , z^\k \}$. 

To close this section let us also add the terms arising from a nontrivial background 
flux $G_4^{\rm flux}$. Combining the gaugings \eqref{MGaugings1andV_flux} with the gauging induced by the non-vanishing $e_{K\L}$ 
one finds 
\beq \label{MGaugings_final}
D q^u = \left\{ \begin{array}{cl} 
d \F +  2 A^\L \theta_\L \, , \quad & \text{ if } q^u = \F \, , \\
d \til \x_K +  A^\L e_{K\L} \,, & \text{ if } q^u = \til \x_K \, , \\
d q^u \, , & \text{ if } q^u \neq \F , \til \x_K \, . 
\end{array} \right.
\eeq
The total potential may then be derived from \eqref{Pcalculation} and \eqref{General5DPotential}. 
The modifications \eqref{MGaugings_final} encode the deviations from a standard Calabi-Yau reduction of M-theory. In the next sections we will demonstrate the up-lift of this five-dimensional gauged supergravity theory to six-dimensions. This will then be interpreted as performing the M-theory to F-theory limit.

\subsection{Circle reduction of gauged 6D supergravity } \label{6D_on_circle}

Having derived the 5D gauged supergravities obtained by M-theory compactifications 
we will now turn to the F-theory side. The starting point will be a general 
6D $(1,0)$ gauged supergravity \cite{Nishino:1997ff, Ferrara:1997gh}. We will dimensionally reduce this theory on 
a circle and then determine the couplings by comparison with the M-theory reduction. 

The 6D theory is specified by a ``pseudo action'' in the sense that self-duality conditions 
for three-form field strengths need to be imposed by hand after variation of the action. 
In the following we will indicate 6D quantities by a $\wh{\phantom{a}}$. The 6D tensor multiplets contain 
a scalar $\wh j^\alpha$ and a two-form $\wh B^\alpha$ with field strength $\wh G^\alpha$ as bosonic degrees of freedom.  
The bosonic fields of the 6D hypermultiplets describe four scalars  $\wh q^U$ each.
The bosonic components of the 6D vector multiplets contain only the vectors $\wh A^I$. These are in general non-Abelian with 
field strength $\wh F^I   = d \wh A^I +\fr12  f^I{}_{J K} \wh A^J \we \wh A^K$.
At lowest order in derivatives the pseudo-action is given by 
\ba
S^{(6)} & = \int_{\cM_6} \bls \fr12 \wh R \wh*1 - \fr14 \wh g_{\a \b} \wh G^\a \we \wh * \wh G^\b - \fr12 \wh g_{\a \b} d \wh j^\a \we \wh * d \wh j^\b - \fr12 \wh h_{UV}  \wh D \wh q^U \we \wh * \wh D \wh q^V \nn \\
&- 2 \O_{\a\b} \wh j^\a b^\b C_{IJ} \wh F^I \we \wh * \wh F^J - \O_{\a \b} b^\a C_{IJ} \wh B^\b \we \wh F^I \we  \wh F^J  - \wh V^{(6)} \wh * \wh 1 \brs \, , 
\label{6DGaugedAction}
\ea
with self-duality condition 
\beq \label{self-duality}
\wh g_{\a\b} \wh * \wh G^\b  = \O_{\a \b} \wh G^\b \, , \quad \text{where}
\qquad \wh G^\a = d \wh B^\a + 2 b^\a \wh \o^{cs} \, , 
\qquad d \wh \o^{cs} = C_{IJ} \wh F^I \we \wh F^J \, .
\eeq
The couplings $b^\alpha,\Omega_{\alpha \beta}$ and $C_{IJ}= \Tr( T_I T_J )$ are constants defining the theory. The
$\wh j^\alpha$ appear in the metric for the tensor multiplets and are normalized as 
\ba
 g_{\a\b} &= 2 \wh j_\a \wh j_\b - \O_{\a \b} \, , &  \wh j^\a \wh j^\b \O_{\a \b} &= 1 \, , &  \wh j_\a &= \O_{\a \b} \wh j^\b \, .
\label{6DIdentities}
\ea
Here $\a = 0, \ldots n^6_T$ is an index in the fundamental of $SO(n^6_T, 1)$ which counts 
the $n^6_T$ tensor multiplets, $I = 1, \ldots \dim (G)$ is an index in the adjoint of $G$ 
which counts the $\dim(G)$ vector multiplets and $U = 1, \ldots, 4 n_H^6$ is an index 
which counts the  $n_H^6$ hypermultiplets. 

As in the 5D case the hypermultiplet gaugings define the covariant derivatives and potential
\ba
\wh D q^U & = d \wh q^U + \wh A^I \wh k_I^U \, , &  \wh V^{(6)} & = - \fr14 \fr{1}{ \O_{\a \b} \wh j^{\a} b^\b} C^{-1 IJ} \wh A_U{}^A{}_B \wh A_V{}^B{}_A \wh k^U_I \wh k^V_J \, , 
\label{6DGaugingDefinitions}
\ea
where $\wh k_I^U$ and $\wh A_U{}^A{}_B$ are in general functions of the hypermultiplet scalars. 
Here $A=1,2$ is an index in the fundamental of the $SU(2)$ R-symmetry of the 6D theory. 
The hypermultiplet gaugings induce a transformation of objects which carry the the 6D R-symmetry 
index such that the covariant derivative of the 6D supersymmetry parameter appearing in the 
gravitino variation is given by
\ba
\wh D_M \wh \e^A & = \wh \na_M \wh \e^A + \wh D_M \wh q^U \wh A_U{}^A{}_B \wh \e^B \, .
\label{6DRSymmetryGauging}
\ea
The covariant derivative of the 6D gravitino $\wh \p_M^A$ appearing in the gauged 
Rarita-Schwinger term also has this structure. 

In order to make contact with the 5D theory we have found in the previous section and 
obtain the F-theory lift we reduce this action on a circle. The ansatz for the 
metric is 
\beq
  \wh s^2_{(6)} = g_{m n} dx^m dx^n + r^2 ( dy - A^0 )^2 \, ,
\eeq
where $A^0$ is the Kaluza-Klein vector, $r$ the circle circumference and $y$ the coordinate along the circle.  
The vector and tensor fields are reduced as 
\beq
 \wh A^I = A^I + \z^I ( dy - A^0 ) \, ,  \qquad \quad 
  \wh B^\a  =B^\a +  (A^\a + 2 b^\a C_{IJ} \z^I A^J) \we ( dy - A^0 ) \, .
\eeq
Substituting this ansatz into the action, integrating over the circle direction, 
performing a Weyl rescaling of the 5D metric $g_{m n} \ra r^{-\fr23} g_{m n}$ and 
using the self duality constraint results in a 5D action with, in general, adjoint 
scalars $\zeta^I$ and non-Abelian vectors $A^I$. We will not display the 
whole non-Abelian action here, since we are mostly interested in the Coulomb branch 
of the theory.


As the 5D M-theory reduction results in the Abelian theories defined by \eqref{MReducedAction3} and \eqref{MReducedAction4}, this must  
be compared with the Coulomb branch of the circle reduced action. The 5D Coulomb branch is 
obtained by giving the adjoint scalars $\zeta^I$ a vacuum expectation value that 
breaks the gauge group as $G \rightarrow U(1)^{\text{rank}(G)}$. We therefore restrict the 
vectors to those which gauge only this Cartan sub-algebra of $G$ which we label as $A^i$ with $i = 1, \ldots, \rank (G)$, for these Cartan elements one has $f^I{}_{i j} = 0$ so that $F^i  = d A^i$.
The scalars in the vector multiplets are accordingly denoted by $\zeta^i$.
To determine the action in the Coulomb branch is in general a hard task, 
since it requires us to integrate out massive fields that gained their mass due to 
the breaking of $G$. In the following we will display the truncated action. 
More precisely, we drop all massive modes that gained their mass 
by moving to the Coulomb branch and the Kaluza-Klein reduction and do
not include corrections arising after integrating out these massive modes.
In principle, one has to compute the Wilsonian effective action 
after integrating out both massive Coulomb branch modes and 
Kaluza-Klein modes \cite{Bonetti:2011mw}. The retained 
fields also include hypermultiplets that admit scalars 
with a gauged 6D shift symmetry. These gaugings will be induced 
by fluxes or the non-Calabi-Yau geometry in the F-theory setup. 
In summary, we will restrict the hypermultiplet scalars to the set $q^u$, where $u = 1, \ldots, n_{H (\Coulomb)}^6$,  which are neutral under the gaugings or 
have only shift symmetries so that $k_i^u = \const$.\footnote{One way of seeing this constraint is to notice that the truncation of the non-Abelian gauge fields $A^I$, which gauge the symmetries of a set of scalars $q^U$, has to be compatible with the equations of motion. On the Coulomb branch we split the vectors into as $A^I = \{A^i, A^{I^\pr} \}$ 
where $A^i$ are the gauge fields associated to the Cartan sub-algebra and $A^{I^\pr}$ are the rest, 
and then set $A^{I^\pr}$ to zero. This is consistent if the $A^{I^\pr}$ field equation $D * F^{I^\pr} = - k^{I^\pr}_U * D q^U + \ldots$ remains satisfied when the truncation is carried out.
Then decomposing the scalars $q^U$ as $q^U = \{ q^u, q^{U^\pr} \}$ where $k^{I^\pr}_u = 0$, $k^{I^\pr}_{U^\pr} \neq 0$ we see 
that when $A^{I^\pr}$ is set to zero  we must 
also set $q^{U^\pr}$ to zero on the right. For this 
reason scalars that are charged under the truncated vectors must also be truncated. However the scalars 
that remain $q^u$ may still be charged under the remaining vectors so that $k^{i}_u \neq 0$ as is seen in our constructions. } These restrictions are made in 
a supersymmetric way so that 
whole multiplets are truncated from the action.

With this restrictions in mind, we are now able to present the 5D action after circle reduction. In order bring the 
action into a more standard form it will be necessary to define the coordinates of the scalar target space \cite{Bonetti:2011mw}
\beq \label{M-def}
M^0 = r^{-\fr43} \, , \qquad \quad
M^\a = r^{\fr23} ( j^\a + 2 b^\a r^{-2} C_{ij} \z^i \z^j ) \, , \qquad \quad
M^i  = r^{-\fr43} \z^i \, . 
\eeq
The action then reads
\ba
S^{(5)}_{(F)} = \int_{\cM_5} \bls & \fr12 R * 1  - \fr12 h_{uv} \, D q^u \we * D q^v - \fr12 G_{\L \S} \, d M^\L \we * d M^\S \nn \\
&- \fr12 G_{\L \S}\, F^{\L} \we * F^\S - \fr1{12} (\cV^{\rm red}_{\L \S \Q}   + X^{\rm red}_{\L \S \Q}) A^\L \we F^\S \we F^\Q - V^{(5)}_{\rm red} * 1\brs \, , 
\label{FReducedAction2}
\ea
where the covariant derivatives for the hypermultiplet scalars are given by $Dq^u = d q^u + A^i k_i^u$. 
The metric $G_{\L \S}$ depends on the scalars $M^\L = (M^0,M^\a,M^i)$ and is given by
\ba 
  G_{\L \S} &= - \fr12 (\pa_{M^\L} \pa_{M^\S} \, \ln \, \cN_{(F)}) |_{\cN_{(F)}=1} \, , 
  &  \cN_{(F)} & \equiv  \cN^{\rm p}_{(F)} + \cN^{\rm np}_{(F)}  \,  ,
  \label{G-6d5d}
\ea
where
\ba \label{DefcNF}
  \cN^{\rm p}_{(F)} & \equiv     \O_{\a \b} M^0 M^\a M^\b - 4 \O_{\a \b} b^\a C_{ij}  M^{\b} M^i M^j \,  ,  &
   \cN^{\rm np}_{(F)} & \equiv    4 \O_{\a\b} b^\a b^\b C_{i j} C_{kl} \fr{ M^i M^j M^k M^l}{M^0} \, . 
\ea 
Let us note that when inserting the definitions \eqref{M-def} into this form of $\cN_{(F)}$ one indeed finds that $\cN_{(F)}=1$ 
as a consequence of $j^\alpha j^\beta \Omega_{\alpha \beta} = 1$. 
The coefficients of the Chern-Simons-type terms are separated into constant couplings $\cV^{\rm red}_{\L \S \Q} $ 
and field-dependent couplings $X^{\rm red}_{\L \S \Theta}(M)$.
The former are given by 
\beq
   \cV^{\rm red}_{\L \S \Q} = \pa_{M^\L} \pa_{M^\S} \pa_{M^\Q}  \, \cN^{\rm p}_{(F)}  \, , 
\eeq
The field dependent Chern-Simons couplings are only 
symmetric in the last two indices $X^{\rm red}_{\L \S \Theta} = X^{\rm red}_{\L (\S \Theta)}$. 
They are given by
\ba
   X^{\rm red}_{0 \L \S}& = X^{\rm red}_{\alpha \L \S} = 0 \ , &  X^{\rm red}_{i \L \S} &= \frac{3}{4}   \pa_{M^i} \pa_{M^\L} \pa_{M^\S}  \, \cN^{\rm np}_{(F)} \ , 
\ea

Finally, let us discuss the scalar potential, by reducing the 6D action we find
\beq
   V^{(5)}_{\rm red} = \fr14 r^{-\fr23}  \fr{1}{ \O_{\a \b} j^{\a} b^\b} C^{-1 ij} A_u{}^A{}_B A_v{}^B{}_A k^u_i k^v_j + \fr12 r^{-\fr83} h_{uv} \z^i \z^j k_i^u k_j^v\ .
\label{RedPotential1}
\eeq
To compare this with the 5D result it is useful to rewrite this expression using the inverse metric $G^{\Sigma \Lambda}$. This requires us to explicitly invert  $G_{\Sigma \Lambda}$ computed 
using \eqref{G-6d5d} and \eqref{DefcNF}. To do this one uses standard inversion formulas for block matrices to find  
\ba
G^{i j} & = \til G^{i j} + \til G \til G^{i k} \til G_k \til G^{j l} \til G_l  = \fr14   r^{-\fr23} \fr1{\O_{\a\b} j^\a b^\b } C^{-1 i j} + 2 r^{-\fr83} \z^i \z^j \, , 
\label{Ginversion}
\ea
where we have applied 
\ba
\til G & =  ( G_{00} - G_{0\a} (G_{\a\b} )^{-1} G_{0 \b} - \til G_{i} \til G^{i j} \til G_j )^{-1}\, , & 
\til G_i & = (G_{0i} - G_{i \a} (G_{\a \b} )^{-1} G_{0 \b} ) \, ,  & \nn \\
\til G^{ij}  &= ( G_{ij} - G_{i \a} (G_{\a\b})^{-1} G_{\b j} )^{-1} \, . 
\ea
and inserted the results for the components of $G_{\L \S}$ which may be read off from \eqref{G-6d5d}. Substituting this into potential \eqref{RedPotential1} we find that this can be rewritten as
\beq \label{red-potential_final}
   V^{(5)}_{\rm red} = - ( ( G^{i j}  - 2 M^i M^j )A_u{}^A{}_B A_v{}^B{}_A k^u_i k^v_j  - \fr12 h_{uv} M^i M^j k_i^u k_j^v ) \, . 
\eeq

\subsection{Lifting to 6D F-theory }\label{Matching5DTheories}

We now wish to match the 5D theory \eqref{FReducedAction2} arising after circle reduction of 6D 
supergravity with the 5D theories \eqref{MReducedAction3} and \eqref{MReducedAction4} in the reduction 
of 11D supergravity. As the actions are both supersymmetric, this can be done by 
matching the hypermultiplet gaugings, the potential and the scalar $\cN$. 

We will first review the matching of $\cN_{(M)}$ given in \eqref{DefcNM} 
with $\cN_{(F)}$ given in \eqref{DefcNF}. To do this we first note that 
the F-theory lift applies in the limit in which the volumes of the elliptic 
fibre and the resolution blowups vanish but where the threefold volume 
remains finite. The effect of taking this limit on the effective action 
results in a rescaling of the scalars as
\ba
M^0 & \ra \e M^0 \, , &  M^\a & \ra \e^{-\fr12} M^\a \, , & M^i  & \ra \e^{\fr14} M^i \, . 
\label{FTheoryLimitRescaling}
\ea
and then taking the limit as $\e \ra 0$. When this is done $\cN_{(M)}$ becomes
\ba
\cN_{(M)} &=  \O_{\a \b} M^0 M^\a M^\b - 4 \O_{\a \b} b^\a C_{ij}  M^{\b} M^i M^j  \, .
\label{restrictedN}
\ea
Next we consider \eqref{DefcNF} this consists of a polynomial and a 
non-polynomial part. The polynomial part $\cN_{(F)}^{\rm p}$ matches \eqref{restrictedN} and the 
non-polynomial part $\cN_{(F)}^{\rm np}$ can be interpreted as a one-loop correction as discussed 
in \cite{Bonetti:2011mw}. Furthermore, $\cN_{(F)}^{\rm np}$ is proportional to 
the contraction $b^\alpha b^\beta \Omega_{\alpha \beta}$ that characterizes the 6D one-loop anomalies.


Let us now discuss the  hypermultiplet gaugings induced by $G_4$-fluxes. 
To do this we compare the gaugings that appear 
in \eqref{MGaugings1andV_flux} with the gaugings \eqref{FReducedAction2}. 
We note from \eqref{FReducedAction2}  that only gaugings associated with the 
vectors $A^i_m$  are present. 
This implies that the $ G_4$-fluxes corresponding to $\q'_0, \q'_\a$ in \eqref{exp-G4flux}, or equivalently 
to the fluxes $\q_0, \q_\a$  defined in \eqref{LtoMredef}, 
cannot be lifted to F-theory. The 6D Killing vectors are related to the remaining 
fluxes and one has
\ba 
     k_i^\F &=  2 \q_i\ , & \q_0 & = \q_\a = 0\ ,
\label{Phi-gauging}
\ea
with all other components of the Killing vectors vanishing. 
It is easy to check that these $k_i^u$  indeed satisfy the Killing vector equations
\ba
\cL_{k_i} h_{uv} = k_i^w \pa_{w} h_{uv} + \pa_{u} k_i^w h_{w v} + \pa_{v} k_i^w h_{w u} =  k_i^\Phi \pa_{\Phi} h_{uv} = 0 \, ,
\ea
as $k_i^v$ are constant and the metric (which can be read off 
from (\ref{MReducedAction2})) is independent of $\F$.

We can see that these allowed gaugings can be lifted to F-theory by considering the reduction 
of Type IIB Supergravity on an orientifold quotient of K3 with D7-Branes which represents the week coupling limit of the F-theory reduction. In this reduction 
the D7 brane action contains a term of the form 
\ba
\int_{D7} \whh C_4 \we \Tr (\whh F \we \whh F) \, , 
\ea
where $\whh{\phantom{a}}$ now indicates a 10D quantity, $\whh F$ is the field strength for the gauge field on the D7 brane and $\whh C_4$ is the IIB Ramond-Ramond 4-form. 
To avoid breaking the 6D Lorentz symmetry of the reduced theory the D7 brane must fill the lower dimensions 
and wrap a 2-cycle $S$ on the internal space. When a flux is turned on such that $\whh F^i  = C^{-1 i j } \q_i [S]$, 
where $[S]$ is the 2-form which is the Poincar\'e dual of $S$, this gives
\ba
\int_{D7} \whh C_4 \we \Tr ( \whh F \we \whh F) = \int_{ \cM_{6} } 2 \wh C_4 \we C_{ij} \wh F^i \int_S C^{-1 j k }  \q_k [S] =  \int_{ \cM_{6} } 2 \wh C_4 \we \wh F^i \q_i \, , 
\label{FTheoryGaugeTerms}
\ea
when the 6D 4-form $\wh C_4$ is dualized to the scalar $\wh \F$ this term is responsible for the appearance
of the gauging in the 6D covariant derivative $D \wh \F  = d \wh \F + 2 \q_i \wh A^i$.  From this we understand that the F-theory dual of the 4-form flux we have described is flux on the world volume of the 7-branes. 

Next we can match the potentials. To do this we simply note that comparing \eqref{General5DPotential} with \eqref{red-potential_final} we find that 
\ba
P_i{}_A{}^B  & = \fr12 k^u_i A_{u}{}^A{}_B \, , & P_0{}_A{}^B & = P_\a{}_A{}^B = 0 \,.  
\ea
Then for the potential induced by the flux gaugings in 5D where \eqref{P3forFlux} applies the 6D potential is given by 
\ba
\wh V^{(6)}_{\flux} =  \fr1{32 \O_{\a\b} \wh j^\a b^\b \wh \cV^2 } C^{-1 i j} \q_i \q_j \, . 
\ea
This potential has  a runaway direction for the scalars $\wh j^\a$ and $\wh \cV$ and as a result the 6D theory effective theory has no maximally symmetric solutions. We will discuss the non-maximally symmetric solution which replace this in the next section.

We can also up-lift the gaugings induced in the reduction on the SU(3) structure manifold. 
As before we compare the gaugings that are arise in the circle reduction \eqref{FReducedAction2} with \eqref{MGaugings2} to find that the only non-vanishing 
killing vectors of the 5D hypermultiplet target space are $k_i^{\til \x_K}  = e_{K i} $ with all other 
components of the killing vectors vanishing. 

We can also consider the F-theory duals of these lifted SU(3) structure deformations. Here we find that the gaugings of the 6D effective theories are caused in the IIB reduction by the presence of extra massive U(1) symmetries. To see this we can note that when these symmetries are included there will be an additional term of the from 
\ba
\int_{D7} \whh C_6 \we \Tr (\whh F)  , 
\ea
where $\whh C_6$ is the Ramond-Ramond 6-form and these extra $U(1)$ branes wrap new cycles $S_i$ on the base $B_2$. To reduce these extra terms  to 6D we expand $ \whh C_6 = \wh Z_{4}^K  \we i_\h \a_K $, where $\h$ is a vector that projects $\a_K$ to a 2-form on the base, and then integrate over $S_i$. This then gives rise to extra terms in the 6D action of the form
\ba
\int_{D7} \whh C_6 \we \Tr (\whh F) = \int_{ \cM_{6} } \wh Z_4^K \we \wh F^i  \int_{S_i} i_\h \a_K = \int_{ \cM_{6} } \wh Z_4^K \we \wh F^i e_{i K} \,. 
\ea
When the 4-form $\wh Z_4^K$ is dualized to give the scalar $\wh {\til \x}_K$ this term then gives rise to gaugings present in our 6D effective theory. We note from this that if we make the gauge choice as described in section \ref{M_on_SU(3)} and expand $\a_K$ into $\a_0$ and $\a_\k$ then, as $i_\h \a_0$ is a $(2,0)$-form and $S_i$ is a $(1,1)$-cycle, we see that $e_{0 i} = 0$ for the F-theory gaugings we describe here. These are then dual to a restricted set of SU(3) structure deformations which also satisfy this constraint. 

As before we can also compare the scalar potentials find that in this case
\ba
\wh V^{(6)}_{U(1)} =  \fr1{32 \O_{\a\b} \wh j^\a b^\b } C^{-1 i j} ( \fr{1}{\cV^2 }e_{\k i} e_{\l j} \x^\k \x^\l + \fr{e^{K_c}}{\cV} e_{\k i} e_{\l j} z^\k \bar z^\l ) \, . 
\label{6DMassivePot}
\ea
When interpreted as coming from D7-branes the potential 
arises by expanding the Dirac-Born-Infeld action. The first term of the potential 
depends on the Wilson line scalars, while the 
second term depends on the D7-brane deformations. The latter 
indicates that certain D7-brane deformations are actually massive
since they require it to wrap a non-supersymmetric cycle.

\section{Vacua and reductions to 4D} \label{sec:6Dvacua}

In this section we will find and comment on certain vacua of the 6D effective theories that result from the F-theory compactifications we have described. In doing this we will approach the effective theories that result from 7-brane fluxes and massive U(1) symmetries separately. In Section \ref{VacuaWithFluxes} we will describe the vacua of the 6D theory deformed by fluxes. As this effective theory has a potential with runaway directions a maximally symmetric solution is not possible and is replaced by vacua which locally describe 4D flat space times a 2D compact internal space. In Section \ref{The 4D Effective Theory} we will consider the 4D effective theories that result from a reduction on the compact 2D part of the solution. In Section \ref{4DMassiveSugras} we will describe the vacua and reductions of the 6D effective theories that result from additional massive U(1) symmetries. 

\subsection{Vacua of 6D F-theory with 7-brane fluxes}
\label{VacuaWithFluxes}

As we have mentioned the 6D gauged supergravity that represents our F-theory reduction with D7-brane flux has no maximally symmetric solution. For this reason it is interesting to investigate what the vacua are. These vacua must solve the 6D equations of motion combined with the pseudo action constraint which are given by 
\ba
\wh R_{M N} & = + \fr14 \wh g_{\a\b} \wh G^\a{}_M{}^{RS} \wh G^\b{}_{N RS} - \fr1{24} \wh  g_{\a\b} \wh G^\a{}^{RST} \wh G^\b{}_{RST} \wh g_{M N}\nn \\
& + 4 \O_{\a\b} \wh j^\a b^\b C_{IJ} \wh F^I{}_M{}^R \wh F^J{}_{N R} - \fr12 \O_{\a\b} \wh j^\a b^\b C_{IJ} \wh F^I{}^{RS} \wh F^J{}_{RS } \wh g_{M N} \nn \\
 & + \wh g_{\a\b} \pa_M \wh j^\a  \pa_N \wh j^\b + \wh h_{UV} \wh D_M \wh q^U \wh D_N \wh q^V + \fr12 \wh V_{(6)} \wh g_{MN}  \, ,  \nn 
\ea
\vspace{-0.7cm}
\ba
d ( \wh h_{UV} \wh * \wh D \wh q^V ) = \fr12 \pa_{U} \wh h_{VW}  \wh D \wh q^V \we \wh * \wh D \wh q^W  + \wh h_{VW} \pa_U \wh k_I^V \wh A^I \we \wh *\wh  D \wh q^W  + \pa_U \wh V_{(6)} \wh * 1 \, , \nn
\ea 
\vspace{-0.7cm}
\ba
d ( \O^{\a \b} \wh g_{\b \g } \wh * d \wh j^\g )  =  \wh j_\b \wh G^\a \we \wh * G^\b  + 2 \wh j_\b d \wh j^\a \we \wh * d \wh j^\b + 2 b^\a C_{IJ}  \wh F^I \we \wh * \wh F^J - \fr1{\O_{\b\g} \wh j^\b b^\g} b^\a \wh V_{(6)}  \wh * 1 \, , \nn
\ea
\vspace{-0.7cm}
\ba 
\wh D( 4 \O_{\a\b} \wh j^\a b^\b  \wh * \wh F^I ) &= - \wh h_{UV} C^{-1 I J } \wh k_{J}^U \wh * \wh D \wh q^V - 4 b^\a \wh g_{\a\b} \wh F^I \we \wh *  \wh G^\b \nn \\& - 2 \O_{\a\b} b^\a b^\b C_{JK} \wh A^I \we  \wh F^J \we \wh F^K  + 4 \O_{\a\b} b^\a b^\b C_{JK} \wh F^I \we \wh \o^{cs} \, , \nn
\ea
\vspace{-0.7cm}
\ba
d ( \O^{\a\b} \wh g_{\b\g}  \wh * \wh G^\g ) &= 2 b^\a  C_{IJ} \wh F^I \we \wh F^J \, , & \wh g_{\a\b} \wh * \wh G^\b & = \O_{\a \b} \wh G^\b \, . 
\ea
This set of equations includes both the fields that correspond to the Coulomb branch, which we have a good understanding of  from the M-theory reduction, as well as the large set of additional degrees of freedom that arise from branes warping shrinking cycles when the F-theory limit is taken. This second set of fields is more mysterious, owing to its non-perturbative origins and consequently we do not know the exact details of the associated couplings. However, when looking for vacua this is not a problem as we know that these additional fields can be consistently truncated out of the theory, leaving only the fields and couplings for which the details are known. For this reason we will only consider vacua which have non-trivial dependence on the Coulomb branch fields. 

In what follows we will be particularly interested in 4D vacua of this 6D theory, we therefore split the 6D world index $M = 0, \ldots 5$ into $\m = 0, \ldots 3$ and $a = 1 , 2$ and we will look for a solution for which the 6D metric is a warped product of 4D Minkowski and some internal space so that 
\ba
d \hat s ^2 = e^{2 W(y^c )} \h_{\m \n } d x^\m dx^\n + g_{a b} ( y^c ) d y^a dy^b \, . 
\ea 
This splitting means that in the vacuum we must have
\ba
\wh G^\a &= 0 \, , & & \text{and} & \wh F^i = \fr12 \fF^i_{a b} dy^a \we dy^b \, , 
\ea
in order to preserve the  4D Lorentz symmetry. The Killing spinor equation then simply reads
\ba
\d \wh \p_M^A & =  \wh \na_M \wh \e^A + \wh D_M \wh q^u \wh A_u{}^A{}_B \wh \e^B  = 0 \, . 
\label{KillingSpinor}
\ea
By considering this equation with the free index pointing in the $\m$ direction we find
\ba
\d \wh \p_\m^A & =   \pa_\m \wh \e^A + \fr14 \wh \o_\m{}_{\n \r} \wh \G^{\n \r} \e^A +  \fr12 \wh \o_\m{}_{\n a} \wh \G^{\n} \wh \G^{a} \wh \e^A +  \fr14 \o_\m{}_{a b} \wh \G^{a b} \wh \e^A + \wh D_m \wh q^u \wh A_u{}^A{}_B \wh \e^B = 0 \, . 
\ea
The term with $\wh \o_{\m \n a}$ cannot cancel anything, so must vanish independently. However, this means that 
\ba
0 = \wh \o_{\m \n a} =  \pa_a W e^{2 W} \h_{\m \n} \, . 
\ea 
So in order for the vacuum to preserve 4D Lorentz invariance we must have $ \pa_a W = 0 $. With an appropriate 4D coordinate redefinition we can then absorb the constant warp factor to give the 6D metric 
\ba
d \hat s ^2 =\h_{\m \n } d x^\m dx^\n + g_{a b} ( y^c ) d y^a dy^b \, , 
\ea 
which we shall consider from now on. The for the $\wh R_{\m \n}$  field equation to be satisfied we then require that in the vacuum
\ba
\O_{\a\b}  \wh j^\a b^\b C_{ij} \wh F^i{}^{a b} \wh F^j_{a b} =  \wh V_{(6)}  \, . 
\label{FluxtoPot}
\ea
Substituting this into the $\wh j^\a$ field equation we find that we may set $d \wh j^\a = 0$ as the runaway direction for $\wh j^\a$ in the potential has been balanced by the $\wh j^\a$ dependence of the flux term. By performing a constant conformal rescaling of the internal space such that $g_{ab} \ra 8 \O_{\a\b} \langle \wh j^\a \rangle b^\b   g_{a b}$ we may then absorb the constant background value of $\wh j^\a$ in all subsequent equations. 

By considering the $\wh q^u$ field equation we find that we can consistently set all the scalars to some constant values apart from $\wh \cV$ which has a runaway direction in the scalar potential and $\wh \F$ which acts as a Stueckelberg field for the gauge potential. The scalars $\wh z^\k$ describe the complex structure moduli of the elliptically fibered Calabi-Yau threefold in our M-theory reduction. Some of these degrees of freedom must therefore also describe the complex structure modulus $\wh \t$ of the auxiliary torus in the F-theory reduction. This means that the kinetic terms for the scalars $\wh z^\k$ may be expanded in terms a kinetic term for $\wh \t$ and kinetic terms for the remaining complex structure moduli as
\ba
- \wh g_{\k \bar \k} d \wh z^k \we * d \wh {\bar z}^{\bar k} = - \fr14 \fr1{ \Im \wh \t^2 } d \wh \t \we * d \wh {\bar \t} + \ldots \, . 
\label{zzExpansion}
\ea
In what follows we will allow $\wh \t$ to vary non-trivially over the 2-dimensional internal space but, to simplify our construction, we will fix the remaining complex structure moduli to be constant.  

To summarize we propose that in our background 
\ba
\wh \cV  &= \fV \, ,& \wh \F & = \ff \, , & \wh \t &= \ft \, , & \wh F^i & = \fF^i \, , 
\ea
where $\fV$ and $\ff$ are real functions of the internal space, $\ft$ is a complex function of the internal space and $\fF^i$ is a real 2-form field strength on the internal space. All other fields of the 6D theory then vanish on the background. 

Using these arguments many of the 6D field equations are solved and the remaining set are greatly simplified giving
\ba
R_{a b}  =  \fr1{2 \Im \ft^2} \pa_{(a} \ft \pa_{b)} \bft   + \fr1{2 \fV^2} \pa_a \fV \pa_b \fV + \fr1{8 \fV^2 } D_a \ff D_b \ff +   \fr{1}{2 \fV^2} C^{-1 i j} \q_i \q_j g_{a b} \, , \nn 
\ea
\vspace{-0.7cm}
\ba
C_{ij} d ( *_2 \fF^j  ) &= - \fr{1}{2 \fV^2} \q_i *_2 D \ff \, , &
C_{ij} \fF^i *_2 \fF^j &= \fr{1}{\fV^2} C^{-1 i j} \q_i \q_j  *_2 1  \, , &
d *_2  d \ft  &=  \fr{-i}{\Im \ft} d \ft \we *_2 d \ft  \nn 
\ea
\vspace{-0.7cm}
\ba
d *_2  d \fV  &= \fr1{\fV} d \fV \we *_2 d \fV - \fr1{4\fV} D \ff \we *_2 D \ff - \fr{1}{\fV} C^{-1 i j} \q_i \q_j  *_2 1 \, ,  &
d ( \fr{1}{\fV^2} *_2 D \ff ) &= 0 \, ,\label{6Dsimplifiedeom}
\ea
where we have used that that $\fF_{a b}^i$ must be proportional to the 2D epsilon tensor as it is a top form on the internal space. 

As the internal space is two-dimensional the 2D Ricci scalar must satisfy $R_{a b} = \fr12 R g_{a b}$. So the R.H.S. of the $R_{a b}$ field equation must also be proportional to $g_{a b}$. We solve this by setting 
\ba
D \ff &= - 2 *_2d \fV \, , &*_2 d \ft &= i d \ft \,, 
\label{FtoV}
\ea
which also solves the $\ff$ and $\ft$ field equations. The two equations for $\fF^i$ are then solved if 
\ba
\fF^i_{a b} = \fr{1}{\fV} C^{-1 ij} \q_j \e_{a b} \, .
\label{Fsolution}
\ea
Furthermore we note that acting with the exterior derivative on \eqref{FtoV} and using \eqref{Fsolution} we recover the equation of motion for $\fV$.

The remaining field equations then describe the geometry of the internal space and the profile of the scalars $\fV$ and $\ft$ on that space. These read
\ba
R & =  - \na^a \na_a  \ln ( \fV \Im \ft )   \, ,  
&\na^a \na_a \fV +   \fr{1}{ \fV} C^{-1 i j} \q_i \q_j  & = 0 \, , & \na^a \na_a \ft  & = 0 \, . 
\label{RVeom}
\ea

To identify the surviving supersymmetry  preserved by this background we can use the Killing spinor equation \eqref{KillingSpinor} with the free index pointing in the $a$ direction. If we assume that 
\ba
\wh \e^A =  e^{ \h(y^a) \e_{ab} \G^{ab} }  \e^A_0 \, , 
\ea
where $\e^A_0$ is a constant spinor, then we find that the Killing spinor equation reads 
\ba
\pa_a \h \e_{b c} \G^{b c} \wh \e^A  + \fr14 \o_{a b c} \wh \G^{ b c } \wh \e^A + D_a \ff \wh A_\F{}^A{}_B \wh \e^B + D_a  \ft  \wh A_\t{}^A{}_B \wh \e^B + D_a \bft  \wh A_{\bar\t}{}^A{}_B \e^B = 0 \, . 
\label{KillingSpinorInternal1}
\ea
As the internal space is two dimensional we can use the identity $ \o_{a b c}  = \o_a \e_{b c} $ to write this expression in terms of $\o_{a}$ and simplify the algebra involved. 
 
We can then make a choice of gamma matrix decomposition where 
\ba
\G^\m & = \g^\m \otimes \g^3 \, , & \G^a & = \id \otimes \g^a \, , 
\ea
where $\g^\m$ are the 4D gamma matrices,  $\g^a$ are the 2D gamma matrices and  $\g^3  = i \g^1 \g^2$. Next we can chose a gauge in which  
\ba
\wh A_\t{}^A{}_B  & = \fr{ i }{ 8 \Im \wh \t} \s^3{}^A{}_B & \wh A_{\bar \t}{}^A{}_B & =  \fr{ i }{ 8 \Im \wh \t} \s^3{}^A{}_B  \, , & \wh A_\F{}^A{}_B & = \fr{ i }{ 8 \wh \cV} \s^3{}^A{}_B \, , 
\label{HyperConnectionCondition}
\ea
If we then impose the constraint 
\ba
\s^3{}^A{}_B \e^B & = \g^3 \e^A \, , 
\label{SUSYbreakingCondition}
\ea
which implies that the background breaks half the supersymmetry of the 6D theory, we find that  \eqref{KillingSpinorInternal1} and \eqref{FtoV} then imply
\ba
\pa_a \h + \fr14 \o_a  & = \fr1{8} \e_{a b} \pa^b \ln ( \fV \Im \ft )  \, . 
\label{OtocV}
\ea
As the 2D Ricci scalar takes the simple form in terms of $\o_a$ 
\ba
R = 2 \e^{a b} \na_a \o_b \, , 
\ea
we find that substituting \eqref{OtocV} into this gives \eqref{RVeom}. So the vacua we have found do indeed break the supersymmetry of the 6D theory by a half. 

Similarly we can look at the killing spinor equation coming from the variation of the vector multiplet fermions. This gives reads
\ba
\wh F^i_{ab} \wh \G^{a b} \wh \e^A +  \fr{ C^{-1\, i j} }{\O_{\a\b} \wh j^\a b^\b}  \wh k_j^{u} \wh A_u{}^A{}_B \wh \e^B = 0 \, . 
\ea
Again substituting  \eqref{Fsolution}  and \eqref{HyperConnectionCondition} into this we find that this equation is satisfied on the constraint \eqref{SUSYbreakingCondition} and so again we find that this background breaks half the supersymmetry of the 6D action. 

We can now consider solutions to \eqref{RVeom}. These may have either constant or varying $\ft$ but must have a non-trivial profile for $\fV$ due to the runaway potential. The solutions with constant $\ft$ correspond to the F-theory lift of 5D domain wall solutions and are described in Appendix \ref{DomainWalls}. However, here we will focus instead on solutions which are dominated by a strongly varying $\ft$ profile. These will correspond to the presence of extra co-dimension 2 sources, for the non-constant $\ft$, in our construction. 

When $\q^i = 0$ and $\fV$ is constant these sorts of solutions are known and are related to cosmic strings \cite{Greene:1989ya}. In this case we may work in a coordinate system where 
\ba
ds^2 &  = \h_{\m \n } d x^\m dx^\n + \fO (z,\bar z)  dz d \bar z \, , 
\ea
in which the self duality condition on $\ft$ becomes
\ba
\pa_{\bar z} \ft  & = 0 \, , & \pa_{ z} \bft & = 0 \, . 
\label{ftSol}
\ea
The solution to the resulting field equations is complicated as there is no known solution with finite energy per unit length for which $\ft$ is both sourced and continuous. Instead the solutions for $\ft$ have discontinuities at which $\ft$ undergoes an $SL(2, \bZ)$ transformation. The solutions are then described by the modular invariant function $j(\ft)$ as 
\ba
j(\ft) = \fr{P(z)}{Q(z)} \, , 
\ea
for polynomials $P$ and $Q$ which share no roots. The roots of these functions then determine the locations and numbers of the co-dimension 2 sources. 

The $z$ dependence of the metric is then determined by the remaining field equation
\ba
\pa_{\bar z} \pa_{ z} \ln \fO &= \pa_{\bar z} \pa_{ z} \ln ( \Im \ft ) \,, 
\label{Om-ftEq}
\ea
which has the modular invariant nowhere vanishing solution
\ba \label{Omega_product}
\fO = \Im \ft | \h (\ft) |^4 \prod_{n = 1}^{N}  \big | (z - z^n)^{-\fr{1}{12} } \big |^2 \, , 
\ea
for $N$ co-dimension 2 sources located at the $z^n$. When $N > 12$ the internal space becomes compact and is given by $\bP^1 $. In this case the only allowed solution has $N = 24$.  As we are interested in compact solutions here this special case will be of particular relevance. 

We now consider turning back on the fluxes $\q^i$. When this is done we modify the metric ansatz so that 
\ba
ds^2 &  = \h_{\m \n } d x^\m dx^\n + \fV ( z ,\bar z) \fO (z,\bar z)  dz d \bar z \,, 
\ea
This ansatz means that the field equations \eqref{ftSol} and \eqref{Om-ftEq} are unmodified when $\fV$ and $\q^i$ are turned on. The remaining field equation for $\fV$ now becomes
\ba
\pa_{\bar z} \pa_{ z} \fV +  C^{-1 i j} \q_i \q_j \fO = 0  \, , 
\label{VEomWithTau}
\ea
solutions to this equation will then describe the geometry of the internal space in the presence of the fluxes $\q^i$ which deform the $\bP^1$ into a new compact space $\wh \cB$.

\subsection{Reduction of the flux deformed effective theory to 4D}
\label{The 4D Effective Theory}

Let us now consider an ansatz for fluctuations about this background for which the internal space is given by $\wh \cB$. As our M-theory analysis gives  only information about the Coulomb branch of the 6D effective theory we will only consider fluctuations in the Coulomb branch fields in our ansatz. This could later be completed to the full set of fields that would be present in the complete F-theory reduction.  

In order to simplify our discussion we will work in the limit where $\q^i$ are small and so we may neglect terms in the reduction which are higher order than $(\q^i) ^2$. The advantage of doing this is that we do not need to explicitly solve the equation \eqref{VEomWithTau} as only structures which are linear in $\fV$ contribute to the effective action. 

We then make an ansatz for the fluctuations where
\ba
d \wh s^2 &= e^{ 2\f } g_{\m \n} d x^\m dx^\n + e^{- 2 \f} ( 1 +  \fr{ C^{-1 i j} \q_i \q_j }{\cV}  ( \D - \P) ) \fO  dz d \bar z \, , \nn 
\ea
\vspace{-0.7cm}
\ba
\wh G^\a & = (d B^\a + 2 b^\a C_{i j} F^i \we A^j) + (d k^\a  +  4  b^\a  \q_i A^i  ) \we *_2 1 \, , &
\wh F^i & =   C^{-1 ij} \q_j *_2 1 +  F^i \, , \nn 
\ea
\vspace{-0.7cm}
\ba
\wh j^\a & = j^\a \, , &
\wh \cV & = \cV ( 1+   C^{-1 i j} \q_i \q_j  \D ) \, , &
\wh \F & = \F ( 1+   C^{-1 i j} \q_i \q_j  \P )  + \ff \,  & 
\label{6DReductionAnsatz}
\ea
where the function $\D$ is related to the background value $\fV$ such that 
\ba
\fV = 1 +  C^{-1 i j} \q_i \q_j  \D \, , 
\ea
so that at the order to which we are working \eqref{VEomWithTau} becomes
\ba
 \pa_{\bar z} \pa_{ z} \D +  \fO = 0  \, , 
\ea
and $\P$ is a constant defined such that  
\ba
\int_{\wh \cB} \P \fO d z \we d \bar z  = \int_{\wh \cB} \D \fO d z \we d \bar z  \, .
\ea
This ansatz satisfies the Bianchi identities for the 6D fields \eqref{6DIdentities} when $F^i = d A^i$. 

In addition to these fluctuations it will be possible to turn on some additional 4D fluctuations in the  6D  hypermultiplet scalars $( \wh \cV, \wh \F, \wh \x^K, \wh {\til \x}_K, \wh z^\k, \wh {\bar z}^\k )$. As the fermions of the 6D theory must be expanded in terms of the constrained background spinor \eqref{SUSYbreakingCondition} we find that only half the hypermultiplet degrees of freedom we can be turned on. For the universal hypermultiplet $( \wh \cV, \wh \F, \wh \x^0, \wh {\til \x}_0 )$ we have already identified that the fluctuations in $\cV$ and $\F$ will be turned on, so fluctuations in $\x^0$ and  $\til \x_0$ are forced to vanish. Alternatively we may divide the remaining hypermultiplets $(
\wh \x^\k, \wh {\til \x}_\k, \wh z^\k, \wh {\bar z}^\k )$ as $\k = \{ \fk, k^\pr \}$ and turn on fluctuations in  $(\x^\fk, \til \x_\fk)$ and $( z^{k^\pr}, \bar z^{k^\pr} )$ such that 
\ba
\wh \x^\fk & = \x^\fk ( 1+ \fr12  C^{-1 i j} \q_i \q_j  \P ) \,, & \wh {\til \x}_\fk & =  {\til \x}_\fk ( 1+  \fr12  C^{-1 i j} \q_i \q_j  \P ) \,, &  \wh z^{k^\pr} & = z^{k^\pr} \, . 
\ea
where $\fk = 1, \ldots ,  n_s$ and $k^\pr = 1, \ldots, h^{1,2} ( \wh Y_3 ) - n_s$. As we will see later supersymmetry then requires that this splitting is performed such that $M_{\fk \fl}$ is an anti-holomorphic function of $z^{k^\pr}$. This can be achieved by performing the split such that $M_{\fk {k^\pr} } = M_{\fk 0 } = 0$ as is shown in \cite{Louis:2002vy, Grimm:2004uq, Louis:2009xd}. 
 
Substituting this into the action and keeping only terms up to and including quadratic order in $\q^i$ we may then reduce the 6D action to 4D. Following this we can impose the self duality condition for $\wh G^\a$ in the standard way and can simplify the action by making the redefinition $\r^\a = e^{-2 \f} j^\a $. The resulting 4D effective theory is then given by
\ba
S^{(4)} & =  \int_{\cM_4} \bls \fr12 R *1  - \fr12 \til g_{\a \b}  D k^\a \we  *  Dk^\b - \fr12 \til g_{\a \b} d \r^\a \we * d \r^\b  - \fr1{ 4 \cV^2} d \cV \we * d \cV \nn \\
& \quad -\fr1{16 \cV^2} (D \F + 2 \x^\fk d \til \x_\fk ) \we *( D \F + 2 \x^\fk d \til \x_\fk )  + \fr{\Im M^{\fk \fl}}{4 \cV} ( d \til \x_\fk - M_{\fk \fm} d \x^\fm) \we * ( d \til \x_\fl - \bar M_{\fl \fn} d \x^\fn )  \nn \\
& \quad - g_{k^\pr \bar k^\pr} d z^{k^\pr} \we * d \bar z^{\bar k^\pr}   - 2 \O_{\a\b} b^\a \r^\b  C_{i j}  F^i  \we  * F^j - 2  \O_{\a \b} b^\a k^\b C_{ij}   F^i \we   F^j  -  V_{\flux}^{(4)}*1\brs \, , 
\label{4DReducedActionGaugings}
\ea
where
\ba
V^{(4)}_{\flux} &= \fr{C^{-1 i j} \q_i \q_j }{32 \r_\a b^\a  } \bigg(   \fr{1}{\cV}   -   \fr{ 8  \r_\a b^\a }{ |\r|^2}  \bigg)^2 \, , &
\til g_{\a\b} & = 2 \fr{\r_\a \r_\b}{|\r|^4 } - \fr{\O_{\a\b}}{|\r|^2} \, , 
\ea
and the gaugings are given by 
\ba
D k^\a &= d k^\a  + 4  b^\a  \q_i A^i  \, , &
D \F & = d \F +2 \q_iA^i \, . 
\ea
The potential appearing here has 3 separate contributions arising from the internal space Ricci tensor, the Kinetic term associated with the non-vanishing 2-form fluxes and the reduction of the 6D potential. These then combine to give the perfect square appearing in the 4D effective theory. 

We note here that the 4D effective theory is not gauge invariant due to the presence of the term $\O_{\a \b} b^\a k^\b C_{ij}   F^i \we F^j $ and the gauged shift symmetry for $k^\a$. This non-invariance is of the sort required to cancel chiral anomalies in the 4D theory and descends from the  equivalent Green-Schwarz mechanism in the 6D action which was required to cancel the anomalies present there \cite{Green:1984sg, RandjbarDaemi:1985wc, Sagnotti:1992qw}. This anomalous variation in the $4D$ theory is crucially related to the flux that has been turned on on $\wh \cB$, as without this the shift symmetry of $k^\a$ is not gauged, so the action is classically invariant.  

In order to make the supersymmetry of this effective theory more apparent we can write the reduced action as
\ba
S^{(4)}  = \int_{\cM_4} \bls \fr12 R *1  - K_{I \bar J} D Y^I \we * D \bar Y^{\bar J}  - 2 \Re ( f )   C_{i j}  F^i  \we  * F^j &\nn \\
  - 2  \Im( f)  C_{ij}   F^i \we   F^j   - \fr1{8 \Re ( f) } C^{- 1 i j} D_i D_j *1   \brs& \, , 
\label{General4DSusyAction}
\ea
where $D Y^I = d Y^I + X^I_i A^i $ and $Y^I$ can be divided into $Y^I = \{ \Tp , T_\a , x_\fk, z^{k^\pr}  \}$. The complex fields given here are related to the real fields appearing in the reduced action by
\ba \label{def-Tx}
\Tp & =   \cV  + i  \fr12 ( \F +  M_{\fk \fl} \x^\fk \x^\fl)  \, , & T_\a & =  \O_{\a\b} (\r^\a  - i k^\a)  \, , & x_\fk  & =  \til \x_\fk - M_{\fk \fl} \x^\fl  \, . 
\ea
The K\"ahler potential and gauge coupling functions are then given  by 
\ba
K &= K(\Tp ,x_\fk) + K(T_\a) + K(z^{k^\pr}) \, , & 
f &=  T_\a b^\b  \, . 
\label{4DKahlerPot} 
\ea
where 
\ba
K(\Tp ,x_\fk) &= - \ln \bigg( \fr12  ( \Tp + \bar \Tp) - \fr18 \Im M^{\fk \fl} (x_\fk - \bar x_\fk)(x_\fl - \bar x_\fl) \bigg) \, , \nn  
\ea
\vspace{-0.7cm}
\ba
K(T_\a) &= - \ln \bigg ( \fr14 \O^{\a\b}  ( T_\a + \bar T_\a ) ( T_\b + \bar T_\b ) \bigg ) \, , & 
K(z^{k^\pr}) & = - \ln \bigg ( \int_{\wh Y^3} \O \we \bar \O \bigg )\, . 
\ea
We note here that for this K\"ahler potential to reproduce the reduced action \eqref{4DReducedActionGaugings} the function of the truncated complex structure moduli $M_{\fk \fl}$ must now be anti-holomorphic in $z^{k^\pr}$ as mentioned above. 

The gaugings are then given by
\ba
X^{\Tp}_i &=  i  \q_i \, , &
X^{T_{\a}}_i  & = - 4 i  b_\a \q_i \, , 
\ea
and the potential may be determined from
\ba 
D_i & =     - \fr{\q_i}{\Tp + \bar \Tp - \fr14 \Im M^{\fk \fl} (x_\fk - \bar x_\fk)(x_\fl - \bar x_\fl) }  +     \fr{ 8 \q_i  (T_\a + \bar T_\a )  b^\a }{ \O^{\a\b} (T_\a + \bar T_\a ) (T_\b + \bar T_\b ) }  \, , 
\ea
which satisfies the standard D-term relation
\ba
i \pa_{\bar I} D_i = K_{ \bar I  J}   X_i^{ J} \, .
\label{DTermConstraint}
\ea

To summarize we note that this reduction gives an N=1 supersymmetric action with the field content listed 
in Table \ref{F-theory_lift1}.

\begin{table}[h!] 
\begin{center}
\begin{tabular}{|c|c|c|} 
\hline  
Complex field & Real components & Index range \\
\hline 
\rule[-.2cm]{0cm}{.7cm} $T_\cB$ & $( \cV, \F)$  &  \\
\hline 
\rule[-.2cm]{0cm}{.7cm} $T_\a$ & $( \r^\a, k^\a )$ & $\a = 1, \ldots, h^{1,1}( B_2 ) $ \\
\hline
\rule[-.2cm]{0cm}{.7cm} $x_\fk$ & $( \til \x_\fk , \x^\fk )$ & $\fk  = 1, \ldots, n_s$ \\
\hline
\rule[-.2cm]{0cm}{.7cm} $z^{k^\pr}$ & & $k^\pr = 1, \ldots,  h^{1,2}( \wh Y_3 ) + n_{SU(3)} - n_s$ \\
\hline
\rule[-.2cm]{0cm}{.7cm} & $A^i_\m$ & $i  = 1, \ldots , h^{1,1}(\wh Y_3) + n_{SU(3)} -  h^{1,1}(B_2) -1$ \\
\hline
\end{tabular} 
\caption{\textit{Fields in the reduction of 6D F-theory on $\wh \cB$.}} \label{F-theory_lift1}
\end{center}
\end{table} 
Here $n_s$ is the number associated with the hypermultiplet splitting in the reduction of the 6D theory and $n_{SU(3)}$ is the number of additional non-harmonic 2-forms introduced when turning on the $SU(3)$ structure deformations in the reduction of M-theory. 

The effective theory has certain gauged shift symmetries that depend on the parameters $\q^i$. Some of these shift symmetries originate from the gauge shift symmetry of the 6D action, while others arise as a result of the fluxes on $\wh \cB$ that must be turned on in the reduction of the 6D theory. In what follows we will show that this action can be interpreted as being a particular limit of F-theory reduced to 4D on a Calabi-Yau fourfold with 7-brane fluxes. 

\subsection{Vacua and reductions of 6D F-theory with massive U(1) symmetries} \label{4DMassiveSugras}

We may also consider the vacua that arise in the 6D F-theory reductions with massive U(1) symmetries. However, here the analysis is significantly simpler. This is because the potential in this case is given by \eqref{6DMassivePot} which is minimised by when 
\ba
e_{\k i} \wh \x^\k =  e_{\k i} \wh z^\k = 0 , 
\ea
This sets the potential to zero in the vacuum so \eqref{FluxtoPot} can be solved with out the need for any fluxes on the 2D internal space to be turned on. The solutions to these theories then simply correspond to the standard cosmic string solutions we have described in Section \ref{VacuaWithFluxes}, with no additional deformation related to the scalar $\fV$, which is now constant. The reduction of the action then proceeds as shown in the previous section but now with $\q_i = 0$. As the massive U(1) gaugings pick out certain 6D hypermultiplet scalars $\wh {\til \x}_{\k}$ which have gauged shift symmetries we find that fluctuations in these scalars must be turned on in the reduction to 4D. 

The action for the 4D effective theory is then given by  
\ba
S^{(4)}  =  \int_{\cM_4} \bls  & \fr12 R *1  - \fr12 \til g_{\a \b}  d k^\a \we  *  dk^\b - \fr12 \til g_{\a \b} d \r^\a \we * d \r^\b  - \fr1{ 4 \cV^2} d \cV \we * d \cV  \nn \\
& - g_{k^\pr \bar k^\pr} d z^{k^\pr} \we * d \bar z^{\bar k^\pr}  -\fr1{16 \cV^2} (d \F + 2 \x^\fk D \til \x_\fk ) \we *( d \F + 2 \x^\fk D \til \x_\fk ) \nn  \\
& + \fr{\Im M^{\fk \fl}}{4 \cV} ( D \til \x_\fk - M_{\fk \fm} d \x^\fm) \we * ( D \til \x_\fl - \bar M_{\fl \fn} d \x^\fn )  \nn \\
&    - 2 \O_{\a\b} b^\a \r^\b  C_{i j}  F^i  \we  * F^j - 2  \O_{\a \b} b^\a k^\b C_{ij}   F^i \we   F^j  -  V^{(4)}_{U(1)}*1\brs \, , 
\ea
where the potential and gaugings are now
\ba
V^{(4)}_{U(1)} & = \fr{C^{-1 i j} e_{\fk i}  e_{\fl j } \x^\fk \x^\fl }{32 \cV^2 \r_\a b^\a  } \, , & D \til \x_\fk & = D \til \x_\fk + e_{\fk i} A^i \, . 
\ea
As before this action can be derived from the standard supersymmetric form \eqref{General4DSusyAction} where the K\"ahler potential and gauge coupling function are given by \eqref{4DKahlerPot}. However, the gaugings and D-Terms are now modified and are instead given by 
\ba
X_{i}^{x_\fk} & = e_{\fk i } \, , & D_i = \fr{e_{\fk i}\fr1{i2} \Im M^{\fk \fl} (x_\fl - \bar x_\fl)  }{ \Tp + \bar \Tp - \fr14 \Im M^{\fk \fl} (x_\fk - \bar x_\fk)(x_\fl - \bar x_\fl)  }\, . 
\ea
which again satisfies \eqref{DTermConstraint}. 

Finally we may consider the effect of turning on both 7-brane fluxes and massive U(1) symmetries simultaneously. The 6D potential is now given by
\ba
\wh V^{(6)} =  \fr1{32 \O_{\a\b} \wh j^\a b^\b } C^{-1 i j} ( \fr{1}{\wh \cV^2 }( \q_i + e_{\k i}  \wh \x^\k) (\q_i + e_{\l j} \wh \x^\l) + \fr{e^{K_c}}{\wh \cV} e_{\k i} e_{\l j} \wh z^\k \wh {\bar z}^\l )\,. 
\label{6DMassivePotwithFluxes}
\ea
To make the vacua that arise from this more apparent it is useful to separate $\q_i$ into those which describe fluxes for the massive U(1) symmetries which we will call $\q_i^{U(1)}$ and those which describe fluxes in the Cartan sub-algebra of the 7-brane field strengths $\q_i^{c}$. These satisfy 
\ba
C^{-1 i j}  \q_i^{c} \q_j^{U(1)} = C^{-1 i j}  \q_i^{c} e_{\k j} & = 0 \, , 
\ea
so the potential may be divided as 
\ba
\wh V^{(6)} =  \fr1{32 \O_{\a\b} \wh j^\a b^\b } C^{-1 i j} ( \fr{1}{\wh \cV^2 } \q^{c}_i  \q^c_i + \fr{1}{\wh \cV^2 }( \q^{U(1)}_i + e_{\k i}  \wh \x^\k) (\q^{U(1)}_i + e_{\l j} \wh \x^\l) + \fr{e^{K_c}}{\wh \cV} e_{\k i} e_{\l j} \wh z^\k \wh {\bar z}^\l ) \, .
\ea
This is then minimised when
\ba
\q_i^{U(1)} + e_{\k i} \wh \x^\k =  e_{\k i} \wh z^\k = 0 , 
\ea
and the resulting vacuum is then simply that described in Section \ref{VacuaWithFluxes} when only the fluxes $\q_i^c$ are turned on. Reducing to 4D as before then gives an effective theory described by the action \eqref{General4DSusyAction} with a K\"ahler potential and gauge coupling function given by \eqref{4DKahlerPot} but now with the gaugings
\ba
X^{\Tp}_i &=  i  \q_i^{U(1)} + i \q^{c}_i  \, , &
X^{T_{\a}}_i  & = - 4 i  b_\a \q^c_i \, , & 
X_{i}^{x_\fk} & = e_{ \fk i } \, , 
\ea
and where the a D-term potential is given by 
\ba
D_i & =   - \fr{\q^c_i + \q_i^{U(1)} -  \fr1{2i} e_{\fk i} \Im M^{\fk \fl} (x_\fl - \bar x_\fl) }{\Tp + \bar \Tp - \fr14 \Im M^{\fk  \fl} (x_\fk - \bar x_\fk)(x_\fl - \bar x_\fl) }  +     \fr{ 8 \q^c_i  (T_\a + \bar T_\a )  b^\a }{ \O^{\a\b} (T_\a + \bar T_\a ) (T_\b + \bar T_\b ) }  \, , 
\ea
This again satisfies \eqref{DTermConstraint} and so gives a supersymmetric action that can be related to F-theory reduced to 4D with 7-brane fluxes and massive U(1) symmetries.

\section{4D F-theory interpretation} \label{sec:4Dsugras}

The vacua of the 6D theory that we identify here can be related to 
vacua of F-theory on a fourfold $\hat Z_4$ which is an elliptic fibration over a 
base $B_3$. The threefold base is chosen to be the direct product 
\beq \label{base_split}
    B_3 = B_2 \times \cB\ ,
\eeq
where $\cB$ is the $\bP^1$ considered in Section \ref{VacuaWithFluxes} before taking into account the back reaction of the flux, which modifies the solution to $\wh \cB$. This threefold base then  admits a K\"ahler structure inherited from $B_2$ and $\cB$.
Furthermore, we propose that the fourfold $\hat Z_4$ with base $B_3$ can also be 
formed by fibering a the threefold $\hat Z_3$ over $\cB$.  As before we will consider 
this threefold to be an elliptic fibration over a base $B_2$. Such a construction is 
well-known for Calabi-Yau fourfolds and threefolds, see for example \cite{Kreuzer:1997zg}, but is expected to 
extend to the more general case considered here. In fact, $\hat Z_3$ and 
$\hat Z_4$ naturally arise as resolutions of singular Calabi-Yau manifolds
in order to be in accord with the interpretation of the massive $U(1)$'s presented 
in Section \ref{Matching5DTheories} \cite{Grimm:2010ez}. 

The solution will then describe two sets of 7-branes, 
a class of 7-branes, including the $\rank( G)$ 7-branes generating the original non-abelian gauge group $G$, 
will wrap cycles on the base $B_2$ as well as wrapping $\cB$ and filling the lower 4-dimensions. 
In addition to this 24 7-branes will wrap the whole of $B_2$ and fill the lower 4-dimensions. 
This number can be determined noting that for a direct product 
\eqref{base_split} one has $c_1(B_3) = c_1(B_2) + c_1(\cB)$. 
By integrating the Kodaira condition over  $\cB$ and 
using that the Euler characteristic of $\cB = \mathbb{P}^1$ is $\c( \cB) = 2$, we see that this matches the known result that 24 sources for $\ft$ are required to form a $\bP^1$ in the cosmic string solution described in Section \ref{VacuaWithFluxes}.   


By construction the reduction of F-theory on the fourfold $\hat Z_4$ gives the 6D effective theory 
we describe here when $\cB$ is very large. However, the intermediate reduction that we have 
described does not capture all the degrees of freedom of the full fourfold reduction. 
In particular certain complex structure moduli $z^n$ associated with the position of the 7-branes on $\cB$ 
in \eqref{Omega_product} are missed. In fact, it is a hard task to fully reconcile the complex structure sector 
of $\hat Z_4$, which is beyond the scope of this work.
Nevertheless, many of the key features of the reduction are captured by our approach and 
we may view this two step analysis as a useful point of view to compute and understand 
certain complicated couplings of the fourfold reduction.
With this interpretation in mind we can understand many of the features of our 6D 
solutions and also link known results in the reductions of F-theory to 4D and 6D. 

A good check that the prescription we have described works in the limit of small fluxes is given by matching the effective theories. Here we simply note that the 4D effective theory we have found by reducing the 6D supergravity in Section \ref{The 4D Effective Theory} matches the 4D effective theory that is given by the reduction of F-theory on a fourfold, with certain complex structure 
moduli and massive 7-brane gauge fields truncated out. 

The first check can be performed in the K\"ahler moduli sector. By constructing  
$B_3$ as in \eqref{base_split} the number 
of degrees of freedom associated to K\"ahler moduli matches. To see this we note that in the 4D theory obtained in Section \ref{The 4D Effective Theory}  the $h^{1,1} (B_2) $  scalars given by $T_\a$,  combine with the one extra scalar given by $\Tp$, to match the $h^{1,1} (B_2) + 1 $ K\"ahler Moduli of the base of the $B_3$. It is therefore useful to label these 
scalars with a combined index $\wh \a = 1, \ldots, h^{1,1} (B_2) + 1 $ such that $T_{\wh \a} = \{ T_\a, T_\cB \}$. 
Furthermore we find that the complex scalars $T_{\wh \a}$ and $x_{\kappa'}$ given in \eqref{def-Tx} 
are then defined in terms of the real variables in a way that matches 
their construction in \cite{Grimm:2004uq}.
Next we note that the volume of $\cB$ is related to the Kaluza Klein scalar $\f$ appearing in our ansatz for the reduction from 6D \eqref{6DReductionAnsatz}  by 
\ba
e^{-2 \f} = \cV_{\cB} \cV^{\fr12} \, , 
\ea
as this takes into account the Weyl rescaling that has been performed to bring the 6D and 4D metrics into the Einstein frame. From this we see that the K\"ahler potential \eqref{4DKahlerPot} can be written as
\ba
K(\Tp, x_\fk) + K(T_\a) = - \ln \cV - \ln( \cV \cV_{\cB}^2 ) = - 2 \ln ( \cV  \cV_{\cB} ) = - 2 \ln( \cV_{B_3} ) \, ,  
\ea
which matches the known result from the reduction of F-theory to 4D.  

Next we can consider the gaugings that are induced by the 7-brane fluxes. The standard result from reductions of F-theory to four dimensions \cite{Grimm:2011fx, Grimm:2011sk} or reductions of M-theory to three dimensions \cite{Haack:2001jz}, is that the K\"ahler moduli receive a gauged shift symmetry described by a matrix $\Q_{\wh \a i } $, which appears in the covariant derivative of $T_{\wh \a} $ as 
\ba
D T_\wh \a = d T_\wh \a - i 4 \Q_{\wh \a i} A^i \, . 
\ea
These $\Q_{\wh \a i} $ are then given in terms of the $G_4$ flux by  
\ba
\Q_{\wh \a i} = \int_{Y_4} G_4 \we \o_{\wh \a} \we \o_i \, . 
\label{4DGaugings}
\ea
Our reduction of the 6D theory involves 7-brane flux which is turned on in two parts. Firstly there is the flux on the $B_2$ in the reduction to the 6D theory and secondly there is the flux on $\cB$ in the reduction to 4D. The total 7-brane flux is then given by 
\ba
\whh F^i =  C^{-1 i j} \q_j  \o_{\cB}  -  C^{-1 i j} \q^j \fr{1}{4 \O_{\b\g} b^\b b^\g } b^\a \o_\a \, . 
\ea
By knowing that the 7-brane flux must be self dual on the 4-cycle on the fourfold wrapped by the 7-branes we can understand the additional term that we have been forced to turn on here as being that which completes the 7-brane flux to a self-dual quantity. 

Then using the standard result that the 7-brane fluxes which we consider here are related to $\whh G_4$ flux in the M-theory dual by $\whh G^{\flux}_4 = \whh F^i \we \o_i $ and substituting this back into \eqref{4DGaugings} we find that the 4D gaugings are given by 
\ba
\Q_{ \cB  i} &=  \int_{\wh Y_4}  - C^{-1 \, j k} \q_j \o_k \we  \fr{1}{\O_{\b\g} b^\b b^\g } b^\a \o_\a \we \o_{\cB} \we \o_i  = - \fr{b^\a C^{-1 \, j k} \q_j }{4 \O_{\b\g} b^\b b^\g }   \int_{\wh Y_3}  \o_i \we \o_k \we  \o_{\a}  \int_{\cB } \o_{\cB}   = \fr14 \q_i \, , &  \nn \\
\Q_{ \a i} &=  \int_{\wh Y_4}   C^{-1 \, j k} \q_j \o_k \we \o_{\cB} \we \o_{\a} \we \o_i  =   C^{-1 \, j k} \q_j  \int_{\wh Y_3}  \o_i \we \o_k  \we  \o_{\a}  \int_{\cB } \o_{\cB}   = -  \q_i b_\a \, , 
\ea
which matches the $X^{T_\a}_i$ and $X^{\Tp}_i$ that we found in Section \ref{The 4D Effective Theory}. From this we see that the action for the K\"ahler moduli that we find by a reduction of the 6D theory matches precisely that found in a direct reduction of F-theory. 

Similarly we can compare the gaugings that are turned on in our reduction of the 6D theory with massive U(1) symmetries with those that are present in the equivalent 4D F-theory reduction. 
Here we see that the scalars $\til \x_\fk$ develop a gauged shift symmetry described by a parameter $e_{i \fk}$ where 
\ba
e_{i k} = \int_{S_i} i_\h \a_\fk \, , 
\ea
which is precisely consistent with the gaugings that are seen due to massive $U(1)$ symmetries which are seen in \cite{Grimm:2011tb}. 

The back-reaction of the $G_4$ flux in the reduction of M-theory to 3D is known to give rise to a warped reduction \cite{Becker:1996gj}. Using the 2-step reduction that we have described we can see that this corresponds to a warped reduction of F-theory. To demonstrate this we may compare the metrics for the 4D and 6D reductions of F-theory that we have described. Firstly in a reduction of IIB to 6D we see that the metric decomposition which leads to an Einstein frame action is given by 
\ba
d \whh s^2 =  \wh \cV^{- \fr12 } \wh g_{MN}d \wh x^M d \wh x^N + g_{i \bar j} dy^i dy^{\bar j} \, , 
\ea
where the factor of $\wh \cV$ in front of the 6D metric gives the required Weyl rescaling in order to cancel the internal space volume factor. When this is further reduced to 4D on the backgrounds we have described this becomes
\ba
d \whh s^2 =  \fV^{- \fr12 }\h_{\m \n} dx^\m dx^\n  +  \fV^{ \fr12 } \fO dz d \bar z + g_{i \bar j} dy^i dy^{\bar j} \, . 
\ea
From which we see that the background value for $\wh \cV$ in the reduction of the 6D action has resulted in an effective warp factor in the reduction to 4D generated by the 6D Weyl rescaling. This relationship can be further emphasised by noting that the 6D field equation for $\fV$ \eqref{VEomWithTau} takes precisely the same form as the warp factor equation in the reduction of M-theory. The observation that in our 6D vacua 7-brane fluxes require a non-trivial profile for $\fV$ then becomes translated to the statement that the flux in the 4D reduction is associated to a non-constant warp factor.

In addition to this we see that, as the potential vanishes in the 6D vacua with only massive U(1) gaugings, there is no need for a flux or a non-trivial profile for $\wh \cV$ in these reductions. This means that no additional component of the 7-brane flux is turned on and no warping is present as a result of these gaugings.

When the 4D F-theory is reduced on a circle, certain one-loop corrections to the 3D Chern-Simons terms are required to match the reduction of the M-theory dual \cite{Grimm:2011fx}. These terms have the form 
\ba
\int_{\cM_3} \Q_{ij} A^i \we F^j 
\ea
where $\Q_{ij}$ is dependent upon the charges and 4D chiralities of the massive tower of 3D fields that have been integrated out. This is given by
\ba
\Q_{ij} = \sum_r q_{ r i} q_{r j} \sum_{\L} \sign ( q_{r \L} v^\L ) \, , 
\ea
where $q_{ r i}$ denotes the charge in the representation $r$ carried by fields. If $\Q_{ij}$ is non-zero then the 4D effective theory is chiral and has associated anomalies. In the reduction of M-theory on $\wh Y_4$ the value of these couplings is related to the $\whh G_4$ flux by
\ba
\Q_{ij} = \int_{\wh Y_4} \whh G_4 \we \o_i \we \o_j \,, 
\ea
Substituting the flux found in our 2-step reduction into this we find that 
\ba
\Q_{ij} =  \q^k \int_{\wh Y_3}  \o_i \we \o_j \we \o_k \int_{\cB} \o_{\cB} =   \cV_{ijk} \q^k \, , 
\label{QtoV}
\ea
where $\cV_{ijk}$ are the intersection numbers which appear in the M-theory reduction on $Y_3$. This indicates that the additional fluxes that are turned on in our intermediate reduction make the 4D effective theory chiral when the additional fields outside of the Coulomb branch are restored. The associated chiral anomalies are then canceled by the Green-Schwarz mechanism \cite{Green:1984sg} referred to in Section \ref{The 4D Effective Theory}.

The constants $\cV_{ijk}$ which appear in this expression are themselves related to one-loop Chern-Simons terms in the circle reduction of 6D F-theory. These are necessary to match with the Chern-Simons terms present in the classical 5D M-theory reduction \eqref{MReducedAction3}. On the F-theory side these may also be expressed in terms of the charges of the fields in the 5D theory as
\ba
\cV_{ i j k } = \sum_r q_{ r i} q_{r j}  q_{r k} \sum_{\L} \sign ( q_{r \L} v^\L ) \, , 
\ea
As in the 3D/4D case, if $\cV_{ijk}$ is non-vanishing then the 6D theory is chiral and may have anomalies which must be canceled. We can then understand \eqref{QtoV} as relating chiral anomalies in 6D and 4D and one-loop Chern-Simons terms in 5D and 3D.

\section{Conclusion}

In this paper we have derived the 6D effective theory resulting 
from a reduction of F-theory on an elliptically fibered
threefold with 7-brane fluxes and massive U(1) symmetries. These effective 
theories were arrived at by considering the reduction of M-theory on an 
elliptically fibered SU(3) structure manifold with $G_4$ flux and making 
use of the duality between M-theory and F-theory. In analysing this 
duality between the 5D and 6D effective theories we see that 7-brane 
fluxes in F-theory are dual to $G_4$ fluxes in M-theory and 
massive U(1) symmetries in F-theory are dual to SU(3) structure 
deformations in M-theory. This agrees with previous discussions 
of the duality between 3D and 4D effective theories carried out in \cite{Denef:2008wq, Grimm:2010ks, Grimm:2011fx, Grimm:2011tb, Grimm:2011sk}. 

The 6D effective theories include hypermultiplets with gauged 
shift symmetries for certain axionic scalars. These gaugings 
result in mass terms for certain 6D vector multiplets and induce 
a potential which may have runaway directions. For 6D effective 
theories which result from turning on 7-brane fluxes the runaway 
direction in the potential means that 6D Minkowski space is no 
longer a solution to the field equations. Instead this solution is 
replaced by a product of 4D Minkowski space and compact 
internal space on which the massive gauge field develops a 
flux. This solution is then similar to that considered in \cite{Salam:1984cj, Aghababaie:2003wz} 
except that now there is a non-trivial profile for an additional 
scalar $\wh \cV$. These solutions break half the supersymmetry 
of the 6D effective theory. 

In addition to this the 6D effective theories may describe a 
non-trivial profile for the complex scalar $\wh \t$ which arises 
as the reduction of the IIB dilaton-axion. If the gauging parameters 
in the 6D effective theory are turned off these solutions become 
those for 6D cosmic strings. Restoring the gauge parameters 
we see that these solutions become modified by the  presence of an additional flux for the massive vector and 
the scalar profile for $\wh \cV$. For small fluxes we were then 
able to find the effective 4D theory that corresponds to the 
reduction on the compact part of the solution. 

As the vacua we have studied break half the supersymmetry of the 6D theory these 4D effective theories have $N=1$ supersymmetry. The nature of this N=2 to N=1 breaking means that only half of the possible modes on the internal space may be supersymmetricly excited and proceeds in a similar way to the orientifold breakings described in \cite{Grimm:2004uq, Louis:2009xd}. As the 4D effective theory is $N=1$ supersymmetric it may now be chiral. This chirality can be confirmed by observing that the effective theory involves Green-Schwarz counter terms which cancel the induced chiral anomalies.  

The 4D effective theories were then related to direct 
F-theory compactifications to 4D with 7-brane fluxes and 
massive $U(1)$ symmetries. Here we found that the additional 
fluxes that must be turned on in the 6D reduction could be 
understood as completing the 7-brane fluxes to a quantity
that is self-dual on the 7-brane internal space. Furthermore 
the non-trivial profile for the scalar $\wh \cV$ can also be
related to a warping in the reduction of F-theory to 4D. This 
analysis shows that many of the complicated effects associated 
with the reduction of F-theory to four dimensions can be captured by the 2-step reduction that we demonstrate here. These 
effects are significantly simpler in the effective theory than in 
their 4D equivalent due to the larger amount of supersymmetry. 

Higher order $\a^\pr$ corrections to the 4D effective theories 
resulting from F-theory compactifications represent a 
challenging problem for F-theory phenomenology.
In further work it would therefore be interesting to investigate 
to what degree these higher order $\a^\pr$ effects in 4D may
be deduced by considering this sort of intermediate reduction. 
These $\a^\pr$ modifications to the 6D effective theory may 
again be easier to deduce as a result of the restrictions due 
to supersymmetry. 

\subsubsection*{Acknowledgements}

We gratefully acknowledge interesting discussions with Federico Bonetti, Denis Klevers, Andre Lukas, 
Raffaele Savelli, Wati Taylor and Kelly Stelle. This research was supported by a grant of the Max Planck Society.

\begin{appendix}
\vspace{2cm} 
\noindent {\bf \LARGE Appendices}

\section{Conventions and Calabi-Yau identities}
\label{Conventions}

In this paper we have used conventions in which the metric in each dimension has a mostly plus signature and 
\ba
\G^\r{}_{\m \n} & = \fr12 g^{\r\s} ( \pa_{\m} g_{\n \s} + \pa_\n g_{\m \s} - \pa_\s g_{\m \n}  ) \, , &
R_{\m \n} & = R^\l{}_{\m \l \n} \, , \nn \\
R^{\l}{}_{\t \m \n} &= \pa_\m \G^\l{}_{\n \t}  - \pa_{\n} \G^\l{}_{\n \t} + \G^\l{}_{\m \s} \G^\s{}_{\n \t} - \G^\l{}_{\n \s} \G^\s{}_{\m \t} \,, &
R & = R_{\m \n} g^{\m \n} \, . 
\ea
We also use conventions in which the $d$ dimensional epsilon tensor  $\e^{\m_1 \ldots \m_d}$ satisfies
\ba
\e_{0 \ldots d-1 } &= \sq{- g} \, , & \e_{\m_1 \ldots \m_n \r_1 \ldots \r_{d-n} } \e^{\n_1 \ldots \n_n \r_1 \ldots \r_{d-n} } & = - n! ( d- n )! \d_{[\m_1}^{\n_1}  \ldots \d_{\m_n]}^{\n_n} \, , 
\ea
for a Lorentzian signature metric. In addition to this we define a p-form $\o_p$ to satisfy
\ba
\o_p & = \fr1{p!} \o_{\m_1 \ldots \m_p} dx^{\m_1} \we \ldots \we dx^{\m_p} \, , \nn \\
d \o_p & = \fr1{p!} \pa_{\n} \o_{\m_1 \ldots \m_p} dx^{\n}  dx^{\m_1} \we \ldots \we dx^{\m_p} \, , \nn \\
* \o_p & = \fr1{p! ( d-p)!} \o_{\m_1 \ldots \m_p} \e^{\m_1 \ldots \m_p}{}_{\n_1 \ldots \n_{d-p} } dx^{\n_1} \we \ldots \we dx^{\n_{d-p}} \, .
\ea

Let us also summarize some useful identities for the complex structure moduli space 
of Calabi-Yau threefolds. The metric on this moduli space is given by 
\beq \label{def-gkappabarkappa}
  g_{\k \bar \k} = - \fr{ \int_{\hat Y_3} \c_\k \we \bar \c_{\bar \k} }{\int_{\hat Y_3} \O \we \bar \O} \, , 
\eeq
where $ \c_\k$ are $(2,1)$ forms on $\hat Y_3$ representing elements of $H^{2,1}(\hat Y_3)$ as
already introduced in \eqref{CYdeformations}. $ g_{\k \bar \k}$ depends through $\c_\k,\O$ on 
the complex structure deformations $z^\k,\bar z^\k$.
One also naturally defines a complex matrix $M_{KL}$ varying over the complex structure moduli space
by setting 
\ba
*_6 \a_K &= A_K{}^L \a_L + B_{KL} \b^L  \, , &
  *_6 \b^K &= C^{KL} \a_L - A_L{}^K \b^L \, ,
\ea
and
\bea \label{def-MKL}
A_K{}^L &=& (\Re M)_{KH} (\Im M)^{-1 HL}  \, , \nn \\ 
 B_{KL} &=& - (\Im M)_{KL} - (\Re M)_{KH} (\Im M)^{-1 HM} (\Re M)_{ML} \, ,  \nn \\ 
C^{KL} &=& (\Im M)^{-1 KL} \, . 
\eea
The imaginary part of $M_{KL}$ is shown to be invertible and here we will denoted this by $\Im M^{KL} = (\text{Im} M )^{-1 KL}$.

\section{6D solutions and 5D domain walls}
\label{DomainWalls}

In Section \ref{VacuaWithFluxes} we studied the vacua of the 6D effective theory that results from F-theory compactified on a Calabi-Yau threefold with 7-brane flux. In that section we were particularly interested in vacua of this effective theory which were dominated by a non-trivial profile for the scalar $\ft$. However, we can instead consider vacua for which $\ft$ is constant. In what follows we will demonstrate that these constant $\ft$ solutions represent the lift of the 5D domain wall solutions that are described in \cite{Lukas:1998tt}. 

To proceed we must chose a coordinate system on the 2D internal space in which to solve  \eqref{RVeom}. Here we will pick this coordinate system such that the results are easy to compare with  \cite{Lukas:1998tt}. To do this we first separate off a circle in the reduction and expand the metric with respect to this while performing a weyl rescaling of the remaining 5D part of the metric such that the 5D action one would arrive at is in the Einstein frame
\ba
ds^2 & = r^{-\fr23} ( a^2 \h_{\m \n } d x^\m dx^\n + b^2 d y^2 ) + r^2 d \f^2 \, , 
\ea
where $r$, $a$ and $b$ are functions of $y$. The solutions of \cite{Lukas:1998tt} satisfy $ b \propto a^4 $ so in order to have an unwarped external space, as required for out solution, we must have 
\ba
a &\propto  r^{\fr13} \, , &   b &\propto r^{\fr43} . 
\ea
Then absorbing the constants of proportionality into the definition of $y$ and $x^\m$ we find that the appropriate coordinate system for carrying out the comparison is
\ba
 ds^2  &  = \h_{\m \n } d x^\m dx^\n + r^2( y )  dy^2 + r^2 (y) d\f^2  \, . 
\label{4D5D6DMetricDecompostion}
\ea
where $y$ is the coordinate normal to the domain wall in the solutions of \cite{Lukas:1998tt}. 

Substituting this into \eqref{RVeom} and requiring that the function $\fV$ depends only on $y$ we find that 
\ba
\pa_y \pa_y \ln{ (r^2) } &= \pa_y \pa_y \ln{ \fV } & 
\pa_y \pa_y \fV +  C^{-1 i j} \q_i \q_j \fr{ r^2}{\fV} & = 0 \, . 
\ea
This is solved by
\ba
\fV & = -  \fr{A}{2}  C^{-1 i j} \q_i \q_j y^2 + B y + C \, , & r^2  & = A \fV \, , 
\label{Vrsolution}
\ea
for some integration constants $A$, $B$, and $C$. 

We can then compare this solution with the results of \cite{Lukas:1998tt} in which 
\ba
a &= \til k \fV^\fr16 \, , &
\fV & = ( \fr16 \cN_{\L \S \Q} f^\L f^\S f^\Q )^2 \, , &
\cN_{\L \S \Q}  f^\S f^\Q & = H_\L = k \q_\L y + k_\L \, , 
\label{LukasSolutions}
\ea
where $\til k$, $k$ and $k_\L$ are constants. As $r \propto a^3$ we find that $ \fV \propto r^2$ which matches our results. However, the solution for $\fV$ does not generally give the quadratic function of $y$ that we find in \eqref{Vrsolution} as is shown in the examples of  \cite{Lukas:1998tt}. This is not surprising as we have seen that a general M-theory reduction cannot be lifted to a 6D F-theory reduction. To restrict to the case where the F-theory lift applies we must first impose that the Calabi-Yau is an elliptic fibration. This means that $\cN$ takes the form shown in \eqref{DefcNM}. Then taking the F-theory limit \eqref{FTheoryLimitRescaling} and using the constraint that $\q_\a = \q_0 = 0$ for the fluxes that can be lifted, we find  
\ba
2 \O_{\a\b} f^\a f^\b &= k_0 \, , & - 16 \O_{\a\b} b^\a f^\b C_{i j} f^j &= H_i \, , & 4 \O_{\a\b} f^\b f^0 - 8 \O_{\a\b} b^\b C_{ij} f^i f^j &= k_\a \, . 
\ea
By contracting these equations in different ways we find that 
\ba
&4 \O_{\a\b} f^\a f^\b f^0 - 8  \O_{\a\b} f^\a b^\b C_{ij} f^i f^j = f^\a k_\a \, , &
4 k_\a f^\a f^0 - 8 k_\a b^\a C_{ij} f^i f^j &= \O^{\a\b} k_\a k_\b \, , & \nn \\
&4 \O_{\a\b} b^\a f^\b f^0 - 8 \O_{\a\b} b^\a b^\b C_{ij} f^i f^j = b^\a k_\a \, , &
- 16 \O_{\a\b} b^\a f^\b C_{i j}  f^i f^j &= f^i H_i \, , & \nn \\
&- 16 \O_{\a\b} b^\a f^\b H_i f^i = C^{-1 i j} H_i H_j \, , & 
2 \O_{\a\b} f^\a f^\b &= k_0 \, . 
\label{fequations}
 \ea
Then imposing that the effective theory is classically gauge invariant requires that $\O_{\a\b} b^\a b^\b = 0$. When this is satisfied  \eqref{fequations} can be rearranged such that substituting back into \eqref{LukasSolutions} gives
 \ba
 \fV & = \left( \fr16  k_0 f^0 + \fr16 H_i f^i + \fr16 k_\a f^\a  \right )^2  = \fr{ ( 8 k_0 k_\a b^\a - C^{-1 i j} H_i H_j ) \O^{\b \g} k_\b k_\g }{256 k_\a b^\a}  \nn \\
 & =  \fr{\O^{\b \g} k_\b k_\g }{256 k_\a b^\a}  ( -  C^{-1 i j} k^2 \q_i \q_j y^2 - 2 C^{-1 i j} k k_i \q_j y  + 8 k_0 k_\a b^\a - C^{-1 i j} k_i k_j   ) \, . 
 \ea 
This solution is once again a quadratic in $y$ and can be matched to our 6D result \eqref{Vrsolution}. In this way we see that the solutions to our 6D field equations with constant $\ft$ can be interpreted as the F-theory lift of the domain wall solutions to the 5D M-theory dual. 

Our solutions \eqref{Vrsolution} may have singular points at the roots of the quadratic where $\fV$ vanishes. The presence of these singularities calls for the introduction of extra sources into the action. To analyse these it is convenient to shift the coordinates $y$ in order to absorb the constant $C$ in  \eqref{Vrsolution}. The solution then has a singularity at $y = 0$ which requires an additional source
\ba
S_{brane} = Q \int_{\cM_{y = 0}} ( \fr1{\wh \cV} \til{*}_4 1 + \til s^* \wh C_4 ) \, , 
\label{Sbrane}
\ea
where Q is a constant to be determined in terms of the integration constants of our solution. In this action $\til *$ is the Hodge dual with respect to the induced metric on the brane source, $\til s^*$ is the pullback to the brane and $\wh C_4$ is a 4-form which is the dual of $\wh \F$ such that $\fr{1}{\wh \cV^2} D \wh \F = 4 *_6  d \wh C_4$.  Our analysis of the gaugings that are introduced by turning on D7-brane flux \eqref{FTheoryGaugeTerms} show that this is descended from the IIB Ramond-Ramond 4-form.

When this source is included the $\fV$ field equation \eqref{RVeom} becomes modified to
\ba
\na^a \na_a \fV +   \fr{1}{ \fV} C^{-1 i j} \q_i \q_j  - 2 Q  \fr{\d(y)}{\sq{g}}  = 0 \, . 
\label{RandVeomwithdelta}
\ea
To solve this we integrate the equation over a Gaussian surface which goes out to a distance  $y = y_0$ away from $y=0$ such that only one singularity is enclosed. After using Stokes law on the total derivative term this gives
\ba
\int_{y = y_0} \sqrt{g_{y_0}} d \f n^a \pa_a \fV + \int_{0\leq y< y_0} \sqrt{g} d y d \f ( \fr{1}{ \fV} C^{-1 i j} \q_i \q_j - 2 Q  \fr{\d(y)}{\sq{g}} ) = 0 \, , 
\label{integratedVeom}
\ea
where $n^a$ is the outward pointing unit normal (satisfying $n_a n^a = 1$) to the surface $y = y_0$ and $ g_{y_0}$ is the determinant of the induced metric on this surface. As we know that the geometry of the solution is given by \eqref{4D5D6DMetricDecompostion} we find that the induced metric and unit normal satisfy
\ba
g_{y_0} &= r^2 \, , & n^y & = \fr1{r} \, , & n^\f & = 0 \, .
\ea
Substituting this into \eqref{integratedVeom} and performing the integration gives
\ba
\pa_y \fV = - A \fr{1}{ \fV} C^{-1 i j} \q_i \q_j  y + 2 Q \, ,
\ea
which matches \eqref{Vrsolution} if $Q = 2 B$. Repeating this argument for the metric field equation or the Bianchi identity for $D \wh \F$ we find again that the source terms \eqref{Sbrane} are required and confirm the relationship between $Q$ and $B$. 

Finally we note that if the 6D theory we have described is reduced on the F-theory circle then the source action \eqref{Sbrane} agrees with the form found in \cite{Khoury:2001wf} for brane sources of 5D domain walls in the M-theory dual. 

\end{appendix}



\end{document}